\documentclass[12pt]{article}

\global\arraycolsep=1pt

\usepackage{amsbsy,amssymb,latexsym,amsfonts, amsmath}
\usepackage{mathrsfs}
\usepackage{graphicx}

\numberwithin{equation}{section}
\newcommand{\bel}[1]{\begin{equation}\label{#1}}                     
\newcommand{\bal}[1]{\begin{eqnarray}\label{#1}}                     
\newcommand{\be}{\begin{equation}}
\newcommand{\ee}{\end{equation}}

\newcommand{\im}{\mathrm{i}}
\newcommand{\ex}{\mathrm{e}}
\newcommand{\de}{\mathrm{d}}
\newcommand{\dis}{\displaystyle}
\newcommand{\scr}{\scriptstyle}
\newcommand{\qq}{\qquad}

\renewcommand{\thefootnote}{\fnsymbol{footnote}}

\newcommand{\bea}{\begin{equation}}
\newcommand{\eea}{\end{equation}}

\begin{document}
%
%
\begin{titlepage}
\begin{flushright}
\normalsize
~~~~
OCU-PHYS 327\\
March, 2010 \\
\end{flushright}

\vspace{15pt}

\begin{center}
{\LARGE
Method of Generating $q$-Expansion Coefficients for} \\
\vspace{5pt}
{\LARGE Conformal Block and $\mathcal{N}=2$ Nekrasov Function}\\
\vspace{5pt}  
 {\LARGE by $\beta$-Deformed Matrix Model }\\
\end{center}

\vspace{23pt}

\begin{center}
{ H. Itoyama$^{a, b}$\footnote{e-mail: itoyama@sci.osaka-cu.ac.jp},
and
T. Oota$^b$\footnote{e-mail: toota@sci.osaka-cu.ac.jp}
}\\
%
\vspace{18pt}
%

$^a$ \it Department of Mathematics and Physics, Graduate School of Science\\
Osaka City University\\
\vspace{5pt}

$^b$ \it Osaka City University Advanced Mathematical Institute (OCAMI)

\vspace{5pt}

3-3-138, Sugimoto, Sumiyoshi-ku, Osaka, 558-8585, Japan \\

\end{center}
%
\vspace{20pt}
\begin{center}
Abstract\\
\end{center}

We observe that, at $\beta$-deformed matrix models for the four-point
conformal block,
the point $q=0$ is the point where the three-Penner type model
becomes a pair of decoupled
two-Penner type models and where, in the planar limit,
(an array of) two-cut eigenvalue distribution(s)
coalesce into (that of) one-cut one(s).
We treat the Dotsenko-Fateev multiple integral, with their paths
under the recent discussion, as perturbed
double-Selberg matrix model (at $q=0$, it becomes a pair of Selberg
integrals)
to construct two kinds of generating functions for
the $q$-expansion coefficients and compute some. A formula associated
with the
Jack polynomial is noted. The second Nekrasov coefficient
for $SU(2)$ with $N_f =4$ is derived.
A pair of Young diagrams appears naturally.
The finite $N$ loop equation at $q=0$ as well
as its planar limit
is solved exactly, providing a useful tool
to evaluate the coefficients as those of the resolvents.
The planar free energy in the $q$-expansion is computed to the lowest
non-trivial order.
A free field representation of the Nekrasov function is given.


\vfill

\setcounter{footnote}{0}
\renewcommand{\thefootnote}{\arabic{footnote}}

\end{titlepage}

\renewcommand{\thefootnote}{\arabic{footnote}}
\setcounter{footnote}{0}

\section{Introduction}
\label{sec:intro}

Matrices have played important roles in the modern developments of
quantum field theory
and string theory. In the last several months,
the $\beta$-deformed quiver matrix models have started serving
as a bridge
\cite{DV,IMO,KMST,EM,SW,Giri,AlM,FHT,Tak,Sulk,Shak,Popo,MMS0911,MMS1001}
that connects the conformal blocks of two-dimensional conformal field
theory \cite{BPZ}
and the low energy effective actions (LEEA) of four-dimensional gauge
theories
having vanishing $\beta$ functions.
In fact, under the conjecture \cite{AGT}, a given Liouville conformal
block gets
identified with the corresponding gauge theory partition function
dominated by the instanton sum.
More than several checks have been provided, supporting
this conjecture and its several extension
\cite{Wyllard,MMM0907,gai0908,MMMM,MM0908a,MM0908b,
NX0908,AGGTV,DGOT,MMM0909a,MMM0909b,pog,MM0909,BT,AY,ABT,
Papado,MM0910,AM0910,Gaiotto0911,NaXi,MM0911}.

These $\beta$-deformed matrix models (or eigenvalue ensembles to call
more properly),
have Virasoro constraints at finite $N$ as Schwinger-Dyson equations
\cite{VirN}.
It is, therefore, not a surprise that the principal part of the
conformal block
is represented by a suitably chosen correlator of the model or
alternatively, a potential, which
has been identified with the Penner-type one \cite{DV,pen}.

Turning our attention to the gauge theory side,
there are two faces of gauge theory LEEA which matrices get connected to.
The one is the effective prepotential captured by
the Seiberg-Witten curve \cite{SeibergWitten}.
It has been demonstrated that the spectral curve of the one-matrix
model \cite{DV}, and more generally that of the
$A_{n}$ quiver matrix model \cite{IMO} are in fact isomorphic to the
corresponding
Seiberg-Witten curve written
in the Witten-Gaiotto form \cite{Witten,Gaiotto}, giving, albeit a
posteriori,
justification of the model.
The other is so-called Nekrasov partition function mentioned above which
arises from the instanton moduli space
and, under the conjecture, is identified with the principal part of the
conformal block.
The expansion coefficients of the former,
the Nekrasov partition function \cite{Nekrasov,NO,NY},
in $q$, the variable that counts the instanton
number, ought to be identified with those of the latter in $q$,
the cross ratio often denoted by $x$
in the literature.

In this paper, we develop computation of these $q$-expansion
coefficients further by three methods
that we present.
The points which we have observed and will elaborate more
in what follows are rather simple.
At $\beta$-deformed matrix models for the four-point conformal block,
the point $q=0$ is the point where the three-Penner type model becomes a
pair of decoupled
two-Penner type models. Information on the $q$-expansion coefficients is
well extracted
at this point by matrix model technology and Selberg type integrals.
In particular, we treat the Dotsenko-Fateev multiple integral \cite{DF}
as perturbed double-Selberg matrix model (at $q=0$, it becomes a pair of
Selberg integrals).
We construct two kinds of generating functions for the $q$-expansion
coefficients,
the one for the conformal block (its principal part)
and the other for the Nekrasov function for $SU(2)$
with $N_f =4$, and compute some.
The second Nekrasov coefficient is derived. A pair of Young diagrams
appears naturally.

In carrying out the computation, the integration domain of the partition
function
has to be introduced properly. In fact, the integration domain of the
partition function of
the conventional Penner type hermitian matrix models is over the entire
eigenvalue coordinates and their
monodromy properties are too simple to deserve a full-fledged four point
conformal block labelled by
five generic anomalous dimensions and must be polished.
The recent work of Mironov, Morozov and
Shakirov \cite{MMS0911,MMS1001} calls attention to the old
Dotsenko-Fateev \cite{DF}
multiple integral and the choice of paths of the screening operators.

In the next section, we consider the Dotsenko-Fateev multiple integral
and treat it as
perturbed double-Selberg matrix model for the expansion in $q$. We note
an integral associated
with the Selberg integral and the Jack polynomial to push our first
method of computation.
In sections, 3 and 4,
we construct the two kinds of generating functions for the $q$-expansion
mentioned above and compute some
coefficients. In particular, the second Nekrasov coefficient is derived.
The 0d-4d dictionary of the parameters is given.
In section 5, computations are carried out from a pair of loop equations
at finite $N$ (our second method)
which are decoupled to each other at $q=0$ and we solve them exactly.
We derive some of the $q$-expansion coefficients again as well as a few
new results.
In section 6, our calculation carried out at the planar level and the
$g_s$ corrections
of the loop equation (our third method) is presented.
Some coefficients of the resolvents and the planar free energy in
$q$-expansion are computed
and are expressed in terms of the $4d$ parameters.
In section 7, a free field representation of the Nekrasov function is given.
In Appendix A, we give a summary of the 0d-4d dictionary, and some
details of section 4 are given.
In Appendix B, we present computation of free energy ($q$-independent
part) in
the $g_s$ expansion from the Selberg integral.

While in this paper we work on the Dotsenko-Fateev multiple integral,
which is a version of the
$\beta$-deformed one-matrix model for $SU(2)$ with $N_f =4$, the point
we have observed,
namely, the decoupling at $q=0$, holds in more general
$\beta$-deformed quiver matrix models
\cite{IMO,MMM,KMMMP,kos1,kos2,CKR,AMOS1,sek,
Hofman,Lazaroiu,KLLR,CT}.
This is seen particularly easily in the planar limit.
In the planar limit, the $SU(n)$ quiver matrix model develops a two-cuts
eigenvalue distribution in the $n=2$ case
and in general an array of $n-1$ two-cuts eigenvalue distributions each
controlled by a multi-Penner potential
of the following type:
\be
c_{1,a} \ln z + c_{2,a} \ln (z-1) + c_{3,a} \ln (z-q), \;\;\; a = 1,
\cdots, n-1.
\ee
The ladder structure can be seen by the form
of the quiver matrix model differential $y_{i}(z)$ \cite{IMO}.
While there are originally two extrema for each of
the potentials, every potential at $q=0$ contains only a single extremum,
indicating that two end-points of two cuts for each
$a$ coalesce to form a one-cut distribution. See the figure for the
$n=3$ case.
The expansion coefficients in $q$ need only to be computed
at an array of one-cut distributions.

\begin{figure}[h]
\begin{minipage}{7cm}
  \begin{center}
  \includegraphics[height=8cm]{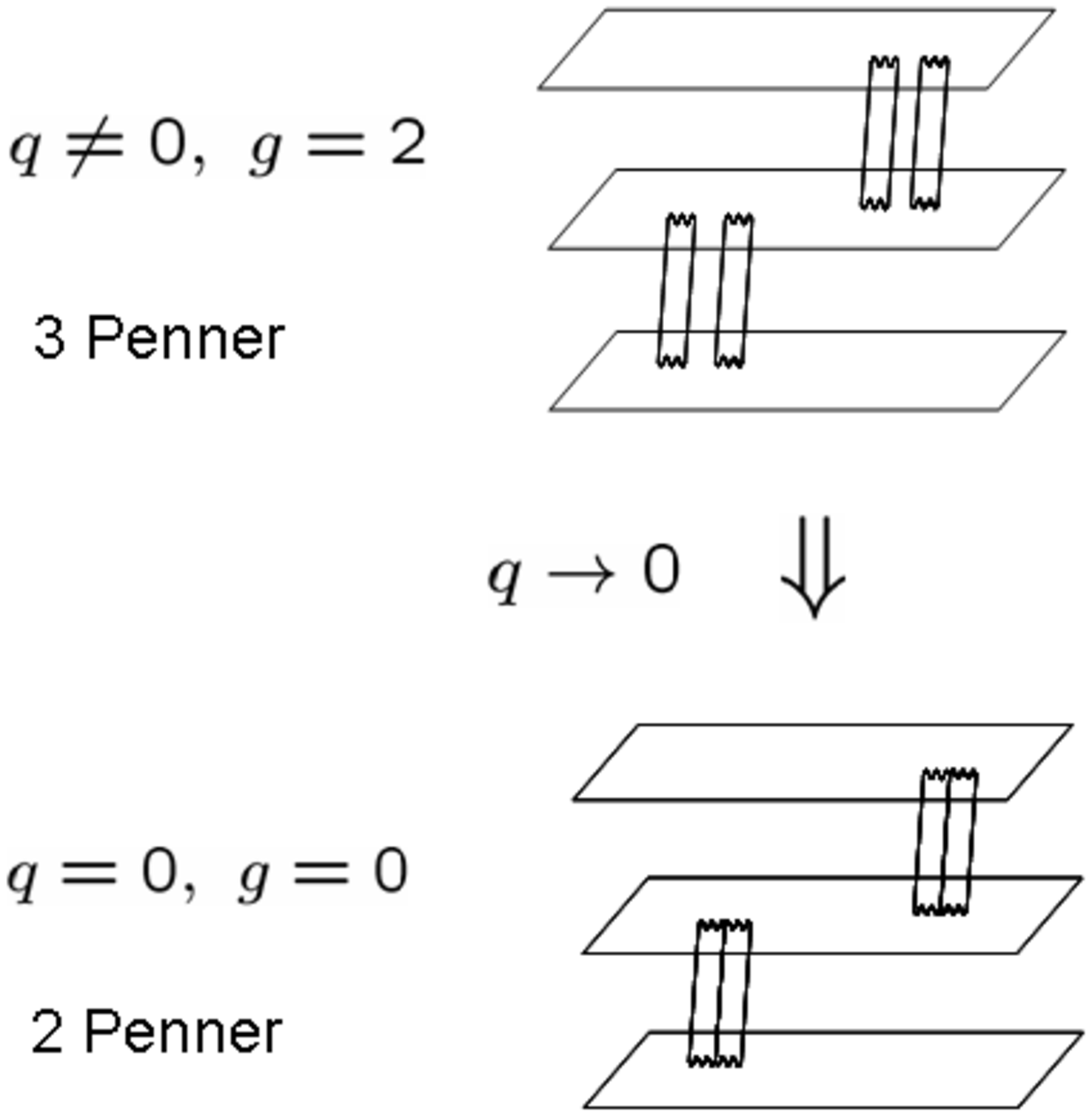} 
  \end{center}
\end{minipage}
\end{figure}


\section{Theory of Perturbed Selberg Integral as Matrix Model}

\subsection{Penner matrix model and
Dotsenko-Fateev multiple integral}
\label{s2-1}

The multi-Penner type ``matrix model'' with
logarithmic potential can be defined by using a 
2d free chiral boson $\phi(z)$ of 2d CFT with $c = 1 - 6 Q_E^2$,
$Q_E = b_E - (1/b_E)$\footnote{
We have changed our notation a little bit from our previous work \cite{IMO}:
$b_E = - \im b$. Normalization of the chiral boson
has been also changed :\ $ \im \sqrt{2} \phi(z) \rightarrow \phi(z)$.
$\langle \phi(z) \phi(w) \rangle = 2 \log(z-w)$.
$T= (1/4) :\partial \phi^2:
+ (Q_E/2) \partial^2 \phi$. With this normalization, the vertex operator
$:\ex^{(1/2) \alpha \phi(z)}:$ has the scaling
dimension $\Delta_{\alpha}=(1/4)\alpha(\alpha-2Q_E)$.}.
The multiple integration originates from the
contour integration of $N$ screening charges.
Problem of Penner-type matrix model is a choice of
integration contours.
They may be closed or may be open.
In \cite{MMS1001}, taking two kinds of open contours
and introducing the ``filling fraction'' $n$ are suggested.
Following  the essence of their suggestion, we introduce two kinds of integration
paths to define
the three-Penner matrix model
\footnote{For simplicity, we consider the CFT on a sphere. Radial ordering
is implicitly assumed in \eqref{Z3P0}.}
:
\bel{Z3P0}
\begin{split}
& Z_{\mathrm{3-Penner}}(q, b_E; N, n, \alpha_1,
\alpha_2, \alpha_3) \cr
&:= \lim_{q_4 \rightarrow \infty}
q_4^{(1/2)(\alpha_4 - 2  Q_E)^2}
\langle 0 |
: \ex^{(1/2) \alpha_1 \phi(0)} :
: \ex^{(1/2) \alpha_2 \phi(q)} :
: \ex^{(1/2) \alpha_3 \phi(1)} :
: \ex^{(1/2) ( \alpha_4 - 2 Q_E) \phi(q_4)} : \cr
& \qq  \qq \qq \qq \times
\left( \int_0^q \de z : \ex^{b_E \phi(z)} : \right)^{n}
\left( \int_1^{q_4} \de z' : \ex^{b_E \phi(z') }: \right)^{N-n}
| 0 \rangle.
\end{split}
\ee
The integer $n$ runs from $0$ to $N$ and serves as
the ``filling fraction''. Our choice of the integration contours $C_1$ and $C_2$
treats the four external points $0$, $q$, $1$ and $\infty$ with
high symmetry: $C_1 = [0,q]$ and 
$C_2=[1,\infty]$\footnote{We tried several choices and judged that 
this choice works best.}.
The constant $\alpha_4$ is determined by
the momentum conservation condition which comes from the zero-mode part:
\be \label{MCC}
\alpha_1 + \alpha_2 + \alpha_3 + \alpha_4 + 2 N b_E =  2 Q_E.
\ee
We will investigate the $q$-expansion of this four-point correlation
function of CFT as a $\beta$-deformed matrix model. Hence, for simplicity
we assume $|q|<1$.

Let us consider the ``intermediate'' channel of \eqref{Z3P0}
at
a point $z$ where $|q|<|z|<1$. The momentum conservation
at this point implies that the internal momentum is uniquely
determined as
\bel{IMM0}
\alpha_I = \alpha_1 + \alpha_2 + 2 n b_E = - \alpha_3 - \alpha_4 - 2(N-n)b_E + 2 Q_E.
\ee
Hence the only allowed intermediate states are $: \ex^{(\alpha_I/2) \phi(z)}:$
and its descendants.
By evaluating the expectation values of the operators,
the partition function \eqref{Z3P0} of the $\beta$-deformed matrix model
turns into the Dotsenko-Fateev multiple integral
\bel{Z3P}
\begin{split}
Z_{\mathrm{3-Penner}} &= q^{(1/2) \alpha_1 \alpha_2}
( 1 - q )^{(1/2)\alpha_2 \alpha_3}
\left( \prod_{I=1}^{n} \int_0^q \de z_I \right)
\left( \prod_{J=n+1}^{N} \int_1^{\infty} \de z_J \right) \cr
& \qq \times
\prod_{I=1}^{n} z_I^{b_E \alpha_1}
(q - z_I)^{b_E \alpha_2} (1 - z_I)^{b_E \alpha_3}
\prod_{1 \leq I < J \leq n} ( z_J - z_I)^{2b_E^2} \cr
& \qq \times
\prod_{J=n+1}^{N} z_J^{b_E \alpha_1}
(z_J - q)^{b_E \alpha_2} (z_J - 1 )^{b_E \alpha_3}
\prod_{n+1 \leq I < J \leq N}
( z_J - z_I)^{2b_E^2} \cr
& \qq \times
\prod_{I=1}^{n} \prod_{J=n+1}^{N}
( z_J - z_I )^{2b_E^2}.
\end{split}
\ee
Therefore, by construction,
we expect that this $\beta$-deformed matrix model
\eqref{Z3P} is also a free field representation of
the conformal block
\be
\mathcal{F}(q\, | \, c\, ; \Delta_1, \Delta_2, \Delta_3, \Delta_4, \Delta_I),
\ee
with $c=1-6Q_E^2$ and\footnote{The vertex operator
$:\ex^{(1/2)( \alpha_4 - 2 Q_E) \phi(q_4)}:$
has the
scaling dimension $(1/4) ( \alpha_4 - 2 Q_E) ( \alpha_4 - 4 Q_E)$,
which is not equal to $\Delta_4 = (1/4) \alpha_4 ( \alpha_4 - 2 Q_E)$.
Even taking into account the overall factor $q_4^{(1/2)(\alpha_4- 2 Q_E)^2}$,
\eqref{Z3P0} does not have expected scaling behaviour at finite $q_4$.
But we send $q_4$ to $\infty$ and in this multi-integral expression \eqref{Z3P},
the parameter $\alpha_4$ does not appear explicitly. The scaling
dimension of \eqref{Z3P} at $\infty$ may be ``dressed'' by the background charge $Q_E$.}
\be
\Delta_i = \frac{1}{4} \alpha_i ( \alpha_i - 2 Q_E), \qq
(i=1,2,3,4), \qq \Delta_I = \frac{1}{4} \alpha_I(\alpha_I- 2Q_E).
\ee

\subsection{perturbed double-Selberg matrix model}

In order to investigate
the $q$-expansion of the multiple integral \eqref{Z3P},
it is convenient to make a change of variables
\be
z_I = q \, x_I, \qq (I=1,2,\dotsc, n), \qq
z_{n + J} = \frac{1}{y_J}, \qq
(J=1,2,\dotsc, N-n).
\ee
The partition function $Z_{\mathrm{3-Penner}}$
\eqref{Z3P} of the three-Penner matrix model
turns into the following form:
\bel{Z3P2}
\begin{split}
Z_{\mathrm{pert}-(\mathrm{Selberg})^2}&= q^{\sigma} (1 - q)^{(1/2)\alpha_2 \alpha_3} \cr
& \times \left(  \prod_{I=1}^{N_L} \int_0^1 \de x_I \right)
\prod_{I=1}^{N_L} x_I^{b_E \alpha_1} ( 1 - x_I)^{b_E \alpha_2}
(1 - q x_I )^{b_E \alpha_3}
\prod_{1 \leq I < J \leq N_L}
| x_I - x_J |^{2b_E^2} \cr
& \times
\left( \prod_{J=1}^{N_R} \int_0^1 \de y_J \right)
\prod_{J=1}^{N_R} y_J^{b_E \alpha_4}
(1 - y_J )^{b_E \alpha_3}
( 1 - q \, y_J )^{b_E \alpha_2}
\prod_{1 \leq I < J \leq N_R}
| y_I - y_J|^{2b_E^2} \cr
& \qq \qq \qq \qq \times
\prod_{I=1}^{N_L} \prod_{J=1}^{N_R} ( 1 - q\, x_I y_J )^{2b_E^2}.
\end{split}
\ee
Here we have renamed $Z_{\mathrm{3-Penner}}$, 
$Z_{\mathrm{pert}-(\mathrm{Selberg})^2}$ and introduced
\be
N_L:= n, \qq N_R:= N-n,
\ee
\bel{sigma}
\sigma:= \frac{1}{2} \alpha_1 \alpha_2 + N_L
+  N_L b_E (\alpha_1 + \alpha_2) + N_L(N_L-1) b_E^2.
\ee
Note that, in this expression, the parameter $\alpha_4$
 has reappeared through the momentum conservation condition \eqref{MCC}:
\bel{MCC2}
\alpha_1 + \alpha_2 + \alpha_3 + \alpha_4 + 2 (N_L + N_R) b_E = 2 Q_E.
\ee
The constrained  set of the parameters $(N_L, \alpha_1, \alpha_2)$ and $(N_R, \alpha_4, \alpha_3)$
enters  \eqref{Z3P2} in a very symmetric way.
Indeed, if we ignore the $q^{\sigma}$ part,
this multiple integral \eqref{Z3P2} is invariant under
the following simultaneous exchange of the parameters:
\bel{parity}
N_L \longleftrightarrow N_R, \qq
\alpha_1 \longleftrightarrow \alpha_4, \qq
\alpha_2 \longleftrightarrow \alpha_3.
\ee
Note that under \eqref{parity}
\bel{sigmad}
\sigma \longleftrightarrow \sigma':=
\frac{1}{2} \alpha_4 \alpha_3 + N_R + N_R b_E ( \alpha_4 + \alpha_3)
+ N_R (N_R - 1 ) b_E^2.
\ee

It is very natural to expect that the related
scaling dimensions to this expected conformal block
$\mathcal{F}(q\,|\,c\,; \Delta_i, \Delta_4,\Delta_I)$
are $\Delta_i=(1/4) \alpha_i(\alpha_i - 2 Q_E)$ $(i=1,2,3)$ and
 $\Delta_4 = (1/4) \alpha_4(\alpha_4 - 2 Q_E)$,
not $(1/4)(\alpha_4 - 2 Q_E)(\alpha_4 - 4 Q_E)$.
In addition to these,
the scaling dimension of the intermediate state is
given by
\be
\Delta_I = \frac{1}{4} \alpha_I ( \alpha_I - 2 Q_E), \qq
\ee
where (recall \eqref{IMM0})
\bel{IMM}
\alpha_I = \alpha_1 + \alpha_2 + 2 N_L b_E = - \alpha_4 - \alpha_4 - 2 N_R b_E
+ 2 Q_E.
\ee
Note that under the symmetry \eqref{parity},
the internal momentum behaves as
$\alpha_I \longleftrightarrow 2 Q_E - \alpha_I$
so $\Delta_I$ is invariant under \eqref{parity}.

We can see that the constant $\sigma$ \eqref{sigma}
and its ``dual'' $\sigma'$ \eqref{sigmad} are equal to
\be
\sigma = \Delta_I - \Delta_1 - \Delta_2, \qq
\sigma' = \Delta_I - \Delta_4 - \Delta_3.
\ee
Hence, the overall $q$ factor of the
partition function \eqref{Z3P2} is consistently identified with that of
the conformal block:
\be
q^{\sigma} = q^{\Delta_I - \Delta_1 - \Delta_2}.
\ee

If we forget the Veneziano factor \cite{LMN,KanMor}
$q^{\sigma}(1-q)^{(1/2)\alpha_2\alpha_3}$,
we  see that at $q=0$ this integral decouples into two independent
Selberg integrals \cite{sel}.
In order to develop its
$q$-expansion, it is more convenient to interpret this
multiple integral as perturbation of two Selberg integrals.
This explains why we have renamed $Z_{\mathrm{3-Penner}}$, 
$Z_{\mathrm{pert}-(\mathrm{Selberg})^2}$ in \eqref{Z3P2}.
The partition function \eqref{Z3P2} of the perturbed double-Selberg model
depends on the cross ratio $q$ and seven parameters:
\be
Z_{\mathrm{pert}-(\mathrm{Selberg})^2}
(q\, | \,  b_E; N_L, \alpha_1, \alpha_2; N_R, \alpha_4, \alpha_3),
\ee
with one constraint \eqref{MCC2}.

We denote the Selberg integral by
\bel{SelI}
\begin{split}
& S_N(\beta_1, \beta_2, \gamma)
= \left( \prod_{I=1}^N \int_0^1 \de x_I \right)
\prod_{I=1}^{N} x_I^{\beta_1 -1} (1-x_I)^{\beta_2-1}
 \prod_{1 \leq I < J \leq N} | x_I - x_J |^{2\gamma}.
\end{split}
\ee
If $N$ is a positive integer and  the complex parameters above obey
\be
\mathrm{Re}\, \beta_1 >0, \ \
\mathrm{Re}\, \beta_2 >0, \ \
\mathrm{Re}\, \gamma > - \mathrm{min}
\left\{
\frac{1}{N},
\frac{\mathrm{Re\, \beta_1}}{N-1},
\frac{\mathrm{Re\, \beta_2}}{N-1}
\right\},
\ee
then the multiple integral \eqref{SelI} is convergent \cite{sel} and
equals to
\bel{SELB}
S_N(\beta_1, \beta_2, \gamma)
= \prod_{j=1}^N
\frac{
\Gamma(1 + j \gamma)
\Gamma(\beta_1 + (j-1) \gamma)
\Gamma(\beta_2 + (j-1) \gamma)}
{ \Gamma(1+\gamma) \Gamma(\beta_1 + \beta_2 + (N+j-2) \gamma)}.
\ee
Hence the perturbed double-Selberg model \eqref{Z3P2}
has a well-defined $q$-expansion if
\be
\mathrm{Re}(b_E \alpha_i) >-1, \qq
(i=1,2,3,4),
\ee
\be
\mathrm{Re}(b_E^2) > - \mathrm{min}
\left\{ \frac{1}{N_L}, \frac{1}{N_R},
\frac{\mathrm{Re}(b_E \alpha_1)+1}{N_L-1},
\frac{\mathrm{Re}(b_E \alpha_2)+1}{N_L-1},
\frac{\mathrm{Re}(b_E \alpha_3)+1}{N_R-1},
\frac{\mathrm{Re}(b_E \alpha_4)+1}{N_R-1}
\right\},
\ee
and $|q|<1$.

Let us denote unperturbed ``Selberg matrix model'' by
\be
\begin{split}
& Z_{(\mathrm{Selberg})^2}(b_E; N_L, \alpha_1, \alpha_2; N_R,
\alpha_4, \alpha_3) \cr
&= Z_{\mathrm{Selberg}}(b_E; N_L, \alpha_1, \alpha_2)
Z_{\mathrm{Selberg}}(b_E; N_R, \alpha_4, \alpha_3) \cr
&:= S_{N_L}(1 + b_E \alpha_1, 1 + b_E \alpha_2, b_E^2)\,
S_{N_R}(1 + b_E \alpha_4, 1 + b_E \alpha_3, b_E^2).
\end{split}
\ee

Averaging with respect to $Z_{(\mathrm{Selberg})^2}$,
$Z_{\mathrm{Selberg}}(N_L)$ and $Z_{\mathrm{Selberg}}(N_R)$
will be denoted by
$\langle\!\langle \dotsm \rangle\!\rangle_{N_L,L_R}$,
$\langle\!\langle \dotsm \rangle\!\rangle_{N_L}$
and
$\langle\!\langle \dotsm \rangle\!\rangle_{N_R}$ respectively.

Then we have an expression of the perturbed double-Selberg model:
\bel{PDS}
\begin{split}
& Z_{\mathrm{pert}-(\mathrm{Selberg})^2}
(q\, | \,  b_E; N_L, \alpha_1, \alpha_2; N_R, \alpha_4, \alpha_3) \cr
&= q^{\sigma} (1-q)^{(1/2) \alpha_2 \alpha_3} \
Z_{(\mathrm{Selberg})^2}(b_E; N_L, \alpha_1, \alpha_2; N_R, \alpha_4,
\alpha_3) \cr
& \qq \times \left\langle \!\!\! \left\langle
\prod_{I=1}^{N_L} (1 - q x_I)^{b_E \alpha_3}
\prod_{J=1}^{N_R} ( 1 - q y_J)^{b_E \alpha_2}
\prod_{I=1}^{N_L}
\prod_{J=1}^{N_R} ( 1 - q x_I y_J)^{2b_E^2}
\right\rangle\!\!\!\right\rangle_{\!\! N_L, N_R}.
\end{split}
\ee
It can be rewritten as
\be
\begin{split}
& Z_{\mathrm{pert}-(\mathrm{Selberg})^2}
(q\, |\, b_E; N_L, \alpha_1, \alpha_2; N_R, \alpha_4, \alpha_3) \cr
&= q^{\Delta_I - \Delta_1 - \Delta_2}\,
\mathcal{B}_0(b_E; N_L, \alpha_1, \alpha_2; N_R, \alpha_4, \alpha_3)\,
\mathcal{B}(q\, |\, b_E; N_L, \alpha_1, \alpha_2; N_R, \alpha_4, \alpha_3),
\end{split}
\ee
where
\be
\mathcal{B}_0(b_E; N_L, \alpha_1, \alpha_2; N_R, \alpha_4, \alpha_3)
:= S_{N_L}(1 + b_E \alpha_1, 1 + b_E \alpha_2, b_E^2)\,
S_{N_R}(1 + b_E \alpha_4, 1 + b_E \alpha_3, b_E^2),
\ee
\bel{calB}
\begin{split}
& \mathcal{B}(q\, |\,  b_E; N_L, \alpha_1, \alpha_2; N_R, \alpha_4, \alpha_3) \cr
&= (1-q)^{(1/2)\alpha_2 \alpha_3}
\left\langle \!\!\! \left\langle
\prod_{I=1}^{N_L} (1 - q x_I)^{b_E \alpha_3}
\prod_{J=1}^{N_R} ( 1 - q y_J)^{b_E \alpha_2}
\prod_{I=1}^{N_L}
\prod_{J=1}^{N_R} ( 1 - q x_I y_J)^{2b_E^2}
\right\rangle\!\!\!\right\rangle_{\!\! N_L, N_R}.
\end{split}
\ee
The function $\mathcal{B}(q) = \mathcal{B}(q\, |\, b_E; N_L, \alpha_1, \alpha_2;
N_R, \alpha_4, \alpha_3)$ has the following form of the $q$-expansion:
\bel{expB}
\mathcal{B}(q) = 1 + \sum_{\ell=1}^{\infty} q^{\ell}\, \mathcal{B}_{\ell}.
\ee

On the other hand, the conformal block of CFT has the form
\be
\begin{split}
& \mathcal{F}(q\,|\, c\, ; \Delta_1, \Delta_2, \Delta_3, \Delta_4, \Delta_I) \cr
&= q^{\Delta_I - \Delta_1 - \Delta_2}\,
\mathcal{B}_0^{(\mathrm{CFT})}
(c\, ; \Delta_1, \Delta_2, \Delta_3, \Delta_4, \Delta_I)
\,
\mathcal{B}^{(\mathrm{CFT})}
(q\,|\, c\,; \Delta_1, \Delta_2, \Delta_3, \Delta_4, \Delta_I),
\end{split}
\ee
where
\be
\mathcal{B}^{(\mathrm{CFT})}(q) = 1 + \sum_{\ell=1}^{\infty} q^{\ell}\,
\mathcal{B}^{(\mathrm{CFT})}_{\ell}.
\ee
The problem is whether the following identity holds or not
\be
\mathcal{B}_0(b_E; N_L, \alpha_1, \alpha_2; N_R, \alpha_4, \alpha_3)
= \mathcal{B}_0^{(\mathrm{CFT})}
(c\, ; \Delta_1, \Delta_2, \Delta_3, \Delta_4, \Delta_I),
\ee
\bel{BB}
\mathcal{B}(q\,|\,  b_E; N_L, \alpha_1, \alpha_2; N_R, \alpha_4, \alpha_3)
= \mathcal{B}^{(\mathrm{CFT})}
(q\,|\, c\,; \Delta_1, \Delta_2, \Delta_3, \Delta_4, \Delta_I),
\ee
if we convert seven constrained parameters
$(b_E; N_L, \alpha_1, \alpha_2; N_R, \alpha_4, \alpha_3)$
into six unconstrained parameters $(c; \Delta_1, \Delta_2, \Delta_3, \Delta_4,
\Delta_I)$ by
\be
c = 1 - 6 \left( b_E - \frac{1}{b_E} \right)^2, \qq
Q_E = b_E - \frac{1}{b_E},
\ee
\bel{0d2dDic}
\Delta_i = \frac{1}{4} \alpha_i ( \alpha_i - 2 Q_E), \qq
(i=1,2,3,4), \qq
\Delta_I = \frac{1}{4} \alpha_I ( \alpha_I - 2 Q_E).
\ee
The internal momentum $\alpha_I$ is given by \eqref{IMM}.
In the next section, we compare $\mathcal{B}_{\ell}$
with $\mathcal{B}_{\ell}^{(\mathrm{CFT})}$ for $\ell=1$.

\subsection{method of calculation: from integrations to combinatorics}

We will be interested in the $q$-expansion of the perturbed double-Selberg model.
It is a special case of more general perturbed Selberg model.
To our surprise, the perturbed Selberg model is
an exactly calculable model.
Let us explain this for more general perturbation.
Consider the following correlation function
\bel{ZpS}
Z_{\mathrm{pert}-\mathrm{Selberg}}(\beta_1, \beta_2, \gamma; \{ g_i \}  )
:=S_N(\beta_1, \beta_2, \gamma)
\left\langle\!\!\!\left\langle
\exp\left( \sum_{I=1}^{N} W(x_I; g )\right)
\right\rangle\!\!\!\right\rangle_{N},
\ee
where the averaging is with respect to the Selberg integral \eqref{SelI}
and
\be
W(x; g )= \sum_{i=0}^{\infty} g_i x^i.
\ee
Suppose that we know the expansion of the exponential of the potential
into the Jack polynomials 
\cite{suth,stan,mac2}\footnote{
 A close connection between the Jack polynomial
and the Virasoro algebra is well-known \cite{mima}.
See also \cite{AMOS1,SSAFR} and references therein.}:
\be
\exp\left( \sum_{I=1}^{N} W(x_I; \{ g_i \} )\right)
= \sum_{\lambda} C_{\lambda}^{(\gamma)}(g) \,
P_{\lambda}^{(1/\gamma)}(x).
\ee
Here $P_{\lambda}^{(1/\gamma)}(x)$ is a polynomial
of $x=(x_1, \dotsm, x_N)$ and  
$\lambda=(\lambda_1, \lambda_2, \dotsm)$ is a partition:
$\lambda_1 \geq \lambda_2 \geq \dotsm \geq 0$. Jack polynomials 
are the eigenstates of
\be
\sum_{I=1}^{N}
\left( x_I \frac{\partial}{\partial x_I} \right)^2
+ \gamma \sum_{1 \leq I <  J \leq N}
\left( \frac{x_I + x_J}{x_I - x_J} \right)
\left(
x_I \frac{\partial}{\partial x_I}
- x_J \frac{\partial}{\partial x_J}
\right),
\ee
with homogeneous degree $|\lambda|=\lambda_1+\lambda_2+
\dotsm$
and are normalized such that for dominance ordering
\be
P_{\lambda}^{(1/\gamma)}(x) = m_{\lambda}(x) + \sum_{\mu < \lambda}
a_{\lambda \mu} m_{\mu}(x).
\ee
Here $m_{\lambda}(x)$ is the monomial symmetric polynomial.

Let $\lambda'$ be the conjugate partition of $\lambda$, i.e., whose
diagram of partition is the transpose of that of $\lambda$ along the main diagonal.
Then Macdonald-Kadell integral \cite{mac1,kad,kan} implies that
\bel{MK}
\begin{split}
\Bigl\langle\!\!\Bigr\langle P_{\lambda}^{(1/\gamma)}(x)
\Bigr\rangle\!\! \Bigr\rangle_{N}
&= \prod_{i \geq 1}
\frac{\dis
\bigl(\, \beta_1+ (N-i) \gamma \, \bigr)_{\lambda_i} \
\bigl(\, (N + 1-i) \gamma \, \bigr)_{\lambda_i}}
{\dis
\bigl(\, \beta_1 + \beta_2 + (2 N-1 - i) \gamma \,
\bigr)_{\lambda_i}} \cr
& \times
\prod_{(i,j) \in \lambda}
\frac{1}{(\lambda_i - j + ( \lambda_j' -i +1 ) \gamma)},
\end{split}
\ee
where $(a)_n$ is the Pochhammer symbol:
\be
(a)_n = a (a+1) \dotsm (a+n-1), \qq
(a)_0 = 1.
\ee
Therefore, the problem of calculating
the correlator \eqref{ZpS} boils down to the problem of
determining  the expansion coefficients $C_{\lambda}^{(\gamma)}(g)$.

\subsection{moments in perturbed double-Selberg model}

Now, we go back to
the perturbed double-Selberg model \eqref{PDS}.
In this case, the parameters in the previous subsection are
specified to $\gamma=b_E^2$ and
\be
N \rightarrow N_L, \qq
\beta_1 \rightarrow 1+b_E \alpha_1, \qq
\beta_2 \rightarrow 1+b_E \alpha_2,
\ee
for the ``left'' part.  We obtain
\bel{MK2}
\begin{split}
\Bigl\langle\!\!\Bigr\langle P_{\lambda}^{(1/b_E^2)}(x)
\Bigr\rangle\!\! \Bigr\rangle_{N_L}
&= \prod_{i \geq 1}
\frac{\dis
\bigl(\, 1 + b_E \alpha_1 + b_E^2 (N_L-i)  \, \bigr)_{\lambda_i} \
\bigl(\, b_E^2(N_L + 1-i)\, \bigr)_{\lambda_i}}
{\dis
\bigl(\, 2 + b_E(\alpha_1 + \alpha_2) + b_E^2 (2 N_L-1 - i) \,
\bigr)_{\lambda_i}} \cr
& \times
\prod_{(i,j) \in \lambda}
\frac{1}{(\lambda_i - j + b_E^2 ( \lambda_j' -i +1 ))}.
\end{split}
\ee
The  expressions for the ``right'' part are obtained by replacing
$x=(x_1,\dotsm, x_{N_L})$ with $y=(y_1,\dotsm, y_{N_R})$
and also by replacing parameters according to \eqref{parity}.

For example,
up to $|\lambda|\leq 2$, explicit forms of the
Jack symmetric polynomials
$P_{\lambda}^{(1/b_E^2)}(x)$ are:
\be
\begin{split}
P_{(1)}^{(1/b_E^2)}(x) &= m_{(1)}(x) = \sum_{I=1}^{N_L} x_I, \cr
P_{(2)}^{(1/b_E^2)}(x) &= m_{(2)}(x)
+ \frac{2b_E^2}{1+b_E^2}\, m_{(1^2)}(x)
= \sum_{I=1}^{N_L} x_I^2 + \frac{2b_E^2}{1+b_E^2} \sum_{1\leq I <J \leq N_L}
x_I x_J, \cr
P_{(1^2)}^{(1/b_E^2)}(x)&= m_{(1^2)}(x) =
\sum_{1\leq I <J \leq N_L}
x_I x_J.
\end{split}
\ee
Hence by using the formula \eqref{MK2} for these cases, we have
\bel{mmt1}
\Bigl\langle\!\!\Bigl\langle b_E P_{(1)}^{(1/b_E^2)}(x)
\Bigr\rangle\!\!\Bigr\rangle_{\! N_L}
= \frac{b_E N_L( b_E N_L - Q_E + \alpha_1)}
{(\alpha_I - 2 Q_E)},
\ee
\bel{p2}
\Bigl\langle\!\!\Bigr\langle
b_E \, P_{(2)}^{(1/b_E^2)}(x)
\Bigr\rangle\!\!\Bigr\rangle_{N_L}
=  \frac{b_E N_L( b_E N_L + b_E^{-1})( \alpha_1 + b_E N_L - b_E + 2 b_E^{-1})}
{(b_E + b_E^{-1})(\alpha_I - 2 b_E + 3 b_E^{-1})(\alpha_I - 2 Q_E)},
\ee
\bel{p11}
2 \,  \Bigl\langle\!\!\Bigl\langle b_E^2 \, P_{(1^2)}^{(1/b_E^2)}(x)
\Bigr\rangle\!\!\Bigr\rangle_{\! N_L}
= \frac{b_E N_L ( b_E N_L - b_E)(\alpha_1 + b_E N_L -Q_E)
(\alpha_1 + b_E N_L - Q_E - b_E)}
{(\alpha_I - 2 Q_E)(\alpha_I - 2 Q_E - b_E)}.
\ee
Here we have used the internal momenta $\alpha_I$ \eqref{IMM}
to simplify expressions.

\section{Matrix Model vs Conformal Block}

In the light of the AGT conjecture, the extensive studies on $q$-expansion of 
the conformal blocks have been carried out in 
\cite{AlM,MMS0911,MMS1001,MMM0907,MMMM,MMM0909b,AM0910}.
Especially, in \cite{MMS0911,MMS1001} the analysis is performed
based on the Dotsenko-Fateev integral. Therefore, we
only consider the $q$-expansion by the $\beta$-deformed matrix model
\eqref{expB} up to the first order.
For higher order expansion coefficients, see these references.

Let us examine the $q$-expansion \eqref{expB}.
Note that $\mathcal{B}(q)$
has the factorized form of the plethystic exponential:
\bel{PleB}
\mathcal{B}(q)
= \left\langle\!\!\!\left\langle
\exp\left[ - 2 \sum_{k=1}^{\infty}
\frac{q^k}{k}
\left( b_E \sum_{I=1}^{N_L} x_I^k + \frac{1}{2} \alpha_2 \right)
\left( b_E \sum_{J=1}^{N_L} y_J^k + \frac{1}{2} \alpha_3 \right)
\right]
\right\rangle\!\!\!\right\rangle_{\!\! N_L, N_R}.
\ee
In the expansion of this plethystic exponential,
a pair of partitions $(Y_1, Y_2)$ naturally appears.
The factorized form explains why in the $\mathcal{B}$-expansion
\be
\mathcal{B}(q) = \sum_{k=0}^{\infty} q^k \sum_{|Y_1|=|Y_2|=k}
\gamma_L(Y_1) \, Q^{-1}(Y_1, Y_2) \, \gamma_R(Y_2),
\ee
such that $\gamma_L(Y_1)$ depends only on the first
partition $Y_1$ and depends only on the ``left'' data
$(b_E, N_L, \alpha_1, \alpha_2)$,
or equivalently  $(c, \Delta_1, \Delta_2, \Delta_I)$.
Similarly
$\gamma_R(Y_2)$ is determined only by the second
partition $Y_2$ and by the ``right'' data $(b_E, N_R, \alpha_4, \alpha_3)$
or $(c, \Delta_4, \Delta_3, \Delta_I)$.

\subsection{$\mathcal{B}_1$}

From \eqref{PleB}, we can read off the first coefficient $\mathcal{B}_1$ as follows
\be
\mathcal{B}_1 = - 2
\left\langle\!\!\!\left\langle b_E \sum_{I=1}^{N_L} x_I + \frac{1}{2} \alpha_2
\right\rangle\!\!\!\right\rangle_{\! \! N_L}
\left\langle\!\!\!\left\langle b_E \sum_{J=1}^{N_R}  y_J + \frac{1}{2} \alpha_3
\right\rangle\!\!\!\right\rangle_{\! \! N_R}.
\ee
Using \eqref{mmt1} and \eqref{0d2dDic}, we find
\be
\begin{split}
\left\langle\!\!\!\left\langle b_E \sum_{I=1}^{N_L} x_I + \frac{1}{2} \alpha_2
\right\rangle\!\!\!\right\rangle_{\! \! N_L}
&= \frac{b_E N_L( \alpha_1 + \alpha_2 + b_E N_L - Q_E)
+ (1/2)\alpha_2 (\alpha_1 + \alpha_2 - 2 Q_E)}
{ \alpha_I - 2 Q_E} \cr
&= \frac{\Delta_I + \Delta_2 - \Delta_1}{(\alpha_I - 2 Q_E)},
\end{split}
\ee
\be
\begin{split}
\left\langle\!\!\!\left\langle b_E \sum_{J=1}^{N_R}  y_J + \frac{1}{2} \alpha_3
\right\rangle\!\!\!\right\rangle_{\! \! N_R}
&= - \frac{b_E N_R( \alpha_4 + \alpha_3 + b_E N_R - Q_E)
+ (1/2)\alpha_3 (\alpha_4 + \alpha_3 - 2 Q_E)}
{ \alpha_I} \cr
&= - \frac{\Delta_I + \Delta_3 - \Delta_4}{\alpha_I}.
\end{split}
\ee
At the first order, the $\beta$-deformed matrix model correctly reproduces the
conformal block of CFT:
\bel{B1OK}
\mathcal{B}_1 =
\frac{(\Delta_I + \Delta_2 - \Delta_1 )( \Delta_I + \Delta_3 - \Delta_4)}{2\Delta_I}
= \mathcal{B}_1^{(\mathrm{CFT})}.
\ee

\section{Matrix Model vs Nekrasov Function}

\label{ChapMMvsNek}

In the previous section, we have considered the $q$-expansion of \eqref{calB}
which contains the Veneziano factor $(1-q)^{(1/2) \alpha_2 \alpha_3}$
as an overall factor. 
Instead, we can consider
another $q$-expansion which does not include this factor.
Let
\bel{BvsA}
\mathcal{B}(q) = (1-q)^{(1/2)\alpha_2 \alpha_3} \mathcal{A}(q),
\ee
where
\bel{calA}
\mathcal{A}(q):=
\left\langle \!\!\! \left\langle
\prod_{I=1}^{N_L} (1 - q x_I)^{b_E \alpha_3}
\prod_{J=1}^{N_R} ( 1 - q y_J)^{b_E \alpha_2}
\prod_{I=1}^{N_L}
\prod_{J=1}^{N_R} ( 1 - q x_I y_J)^{2b_E^2}
\right\rangle\!\!\!\right\rangle_{\!\! N_L, N_R}.
\ee
It also has the $q$-expansion
\be
\mathcal{A}(q) = 1 + \sum_{\ell=1}^{\infty} q^{\ell}\, \mathcal{A}_{\ell}.
\ee
This expansion is essentially the Nekrasov partition function.
The analysis of the expansion of the Nekrasov function
can be found in \cite{MM0908a,MM0908b,MM0911,Sulk}.

\subsection{0d-4d relation for the perturbed double-Selberg model}

Before going on to examine the expansion of \eqref{calA}, we determine
the 0d-4d dictionary for the $\beta$-deformed matrix model.
In order to obtain this,  Eq. \eqref{BvsA} at the first order in $q$
gives strong restrictions on parameter relations. The first order relation of 
\eqref{BvsA} is
\bel{BvsA1}
\mathcal{B}_1 = \mathcal{A}_1 - \frac{1}{2} \alpha_2 \alpha_3.
\ee
As an input, we use the explicit form of the Nekrasov function:
\be
\mathcal{A}_1^{\mathrm{Nek}} = 
\mathcal{A}_{(1),(0)}^{\mathrm{Nek}} + 
\mathcal{A}_{(0),(1)}^{\mathrm{Nek}},
\ee
where
\bel{NekA10}
\mathcal{A}_{(1),(0)}^{\mathrm{Nek}}
= \frac{(a+m_1)(a+m_2)(a+m_3)(a+m_4)}{2a(2a+\epsilon) g_s^2},
\ee
\bel{NekA01}
\mathcal{A}_{(0),(1)}^{\mathrm{Nek}}
= \frac{(a-m_1)(a-m_2)(a-m_3)(a-m_4)}{2a(2a-\epsilon) g_s^2}.
\ee
The numerator of the left-handed side of \eqref{BvsA1}
should be factorized into the product of
the left-part data $\Delta_I + \Delta_2 - \Delta_1$
and the right-part data $\Delta_I + \Delta_3 - \Delta_2$.
This factorization can be explained only by the following
identity:
\be
\begin{split}
& \frac{\prod_i ( a + m_i)}{2a(2a+\epsilon) g_s^2}
+ \frac{\prod_i ( a - m_i)}{2a(2a-\epsilon) g_s^2} 
- \frac{1}{2 g_s^2} (m_1+m_2)(m_3+m_4) \cr
&= 
\frac{2}{(2a+\epsilon)(2a-\epsilon) g_s^2}
\left[ a^2 - \frac{1}{2} \epsilon (m_1+m_2) + m_1 m_2 \right]
\left[ a^2 - \frac{1}{2} \epsilon (m_3+m_4) + m_3 m_4 \right].
\end{split}
\ee
By comparing this relation with
\be
\mathcal{A}_1 - \frac{1}{2} \alpha_2 \alpha_3
= \frac{(\Delta_I + \Delta_2 - \Delta_1)( \Delta_I + \Delta_3 - \Delta_4)}
{2 \Delta_I},
\ee
we find natural identification
\be
\alpha_2 = \frac{1}{g_s} ( m_1 + m_2), \qq
\alpha_3 = \frac{1}{g_s} ( m_3 + m_4),
\ee
and proportionality relations
\be
(2a + \epsilon) ( 2 a - \epsilon) \propto \Delta_I,
\ee
\be
a^2 - \frac{1}{2} \epsilon (m_1+m_2) + m_1 m_2
\propto \Delta_I + \Delta_2 - \Delta_1,
\ee
\be
a^2 - \frac{1}{2} \epsilon (m_3 + m_4) + m_3 m_4 
\propto \Delta_I + \Delta_3 - \Delta_4.
\ee
Since $\Delta_2 = (1/4) \alpha_2 ( \alpha_2 - 2 Q_E)$
and $\Delta_3 = (1/4) \alpha_3 ( \alpha_3 - 2 Q_E)$,
we can determine the 0d-4d relation for the $\beta$-deformed 
matrix model as follows:
\begin{align} \label{0d4d}
b_E N_L &= \frac{a-m_2}{g_s},&
b_E N_R &= - \frac{a+m_3}{g_s}, \cr 
\alpha_1 &= \frac{1}{g_s}( m_2 - m_1 + \epsilon ),&
\alpha_2 &= \frac{1}{g_s}( m_2 + m_1 ), \cr
\alpha_3 &= \frac{1}{g_s}( m_3
 + m_4 ),&
\alpha_4 &= \frac{1}{g_s}( m_3 - m_4 + \epsilon ).
\end{align}
Also $b_E = \epsilon_1/g_s$ and $\epsilon= \epsilon_1 + \epsilon_2$,
$(1/b_E) = - \epsilon_2/g_s$.
These relations convert the seven constrained parameters of the matrix model
\be
b_E, \ \  N_L, \ \ \alpha_1, \ \ \alpha_2, \ \ 
N_R, \ \ \alpha_4, \ \ \alpha_3
\ee
into the six unconstrained parameters of the $\mathcal{N}=2$ $SU(2)$ gauge
theory with $N_f=4$:
\bel{GTP}
\frac{\epsilon_1}{g_s}, \ \ 
\frac{a}{g_s}, \ \ 
\frac{m_1}{g_s}, \ \ 
\frac{m_2}{g_s}, \ \ 
\frac{m_3}{g_s}, \ \
\frac{m_4}{g_s}.
\ee
Here $a$ is the vacuum expectation value of the adjoint scalar,
$m_i$ are mass parameters and $\epsilon_1$ is one of the Nekrasov's
deformation parameter.
With these parameters \eqref{GTP}, the momentum conservation condition \eqref{MCC2} 
is automatically satisfied and gives no restriction on them.
Under the exchange of parameters \eqref{parity}, the parameters of the gauge theory
behave as follows:
\bel{GTparity}
a \longleftrightarrow - a, \qq
m_1 \longleftrightarrow m_4, \qq
m_2 \longleftrightarrow m_3.
\ee
Other parameters $g_s$, $\epsilon_1$, $\epsilon_2$, $\epsilon$ are invariant.

Using the 0d-4d relation, we can rewrite the formula \eqref{MK2} in the 
parameters of the gauge theory.
For example, \eqref{mmt1}, \eqref{p2} and \eqref{p11} are rewritten as follows:
\bel{JackMMT}
\begin{split}
\left\langle\!\! \left \langle
b_E P_{(1)}^{(1/b_E^2)}(x) 
\right\rangle\!\! \right\rangle_{\! N_L}
&= \frac{(a-m_1)(a-m_2)}
{g_s (2a - \epsilon)}, \cr
\left\langle\!\! \left \langle
b_E P_{(2)}^{(1/b_E^2)}(x) 
\right\rangle\!\! \right\rangle_{\! N_L}
&= \frac{(a-m_1)(a-m_2)(a-m_1 - \epsilon_2)(a-m_2-\epsilon_2)}
{g_s ( \epsilon_1 - \epsilon_2)(2a - \epsilon)( 2a - \epsilon - \epsilon_2)}, \cr 
\left\langle\!\! \left \langle
2 b_E^2  P_{(1^2)}^{(1/b_E^2)}(x) 
\right\rangle\!\! \right\rangle_{\! N_L}
&= \frac{(a-m_1)(a-m_2)(a-m_1 - \epsilon_1)(a-m_2-\epsilon_1)}
{g_s^2 (2a - \epsilon) ( 2a - \epsilon - \epsilon_1)}.
\end{split}
\ee
In general, we find
\be
\begin{split}
& \Bigl\langle\!\!\Bigl\langle P_{\lambda}^{(1/b_E^2)}(x)
\Bigr\rangle\!\!\Bigr\rangle_{\! N_L} \cr
&= \prod_{(i,j) \in \lambda}
\frac{(-a+m_1 + \epsilon_1 (i-1) + \epsilon_2(j-1)\, )
( - a + m_2 + \epsilon_1(i-1) + \epsilon_2(j-1)\, )}
{(2a - \epsilon - \epsilon_1 ( i-1) - \epsilon_2(j-1) \,)
\, (\, \epsilon_1(\lambda_j' - i +1) - \epsilon_2(\lambda_i - j ))},
\end{split}
\ee
and likewise
\bel{GTPL}
\begin{split}
& \Bigl\langle\!\!\Bigl\langle P_{\lambda}^{(1/b_E^2)}(y)
\Bigr\rangle\!\!\Bigr\rangle_{\! N_R} \cr
&= (-1)^{|\lambda|} \prod_{(i,j) \in \lambda}
\frac{(a+m_3 + \epsilon_1 (i-1) + \epsilon_2(j-1)\, )\, 
( a + m_4 + \epsilon_1(i-1) + \epsilon_2(j-1)\, )}
{(2a + \epsilon + \epsilon_1 ( i-1) + \epsilon_2(j-1) \,)
\, (\, \epsilon_1(\lambda_j' - i +1) - \epsilon_2(\lambda_i - j ))}.
\end{split}
\ee

\subsection{expansion of $\mathcal{A}(q)$}

Now let us consider the expansion of \eqref{calA}.
Note that $\mathcal{A}(q)$ \eqref{calA}
takes the form of the plethystic exponential
given by the sum of two factorized terms:
\bel{PleA}
\begin{split}
\mathcal{A}(q)
&= \left\langle\!\!\!\left\langle
\exp
\left[
- \sum_{k=1}^{\infty} \frac{q^k}{k}
\left( \alpha_2 + b_E \sum_{I=1}^{N_L} x_I^k \right)
\left( b_E \sum_{J=1}^{N_R} y_J^k \right)  
\right. \right. \right. \cr
& \qq \qq \left. \left. \left.
- \sum_{k=1}^{\infty} \frac{q^k}{k}
\left( b_E \sum_{I=1}^{N_L} x_I^k \right)
\left( \alpha_3 + b_E \sum_{J=1}^{N_R} y_J^k \right)
\right]
\right\rangle\!\!\!\right\rangle_{\!\! N_L, N_R}.
\end{split}
\ee
Also in this case a pair of partitions $(Y_1, Y_2)$
appears in the expansion but now $\mathcal{A}(q)$
has the form
\be
\mathcal{A}(q) = \sum_{k=0}^{\infty} q^k
\sum_{|Y_1|+ |Y_2|=k} \mathcal{A}_{Y_1, Y_2}.
\ee
Here the expansion coefficients depend on both partitions
$(Y_1,Y_2)$ and depends on all data
$(a, m_1, m_2, m_3, m_4, g_s, \epsilon_1)$.

For example, at the first order we have
\bel{A1-0}
\begin{split}
& \mathcal{A}_1 \cr
&=-
\left\langle\!\!\!\left\langle
\alpha_2 + b_E \sum_{I=1}^{N_L} x_I
\right\rangle\!\!\!\right\rangle_{\!\! N_L}
\left\langle\!\!\!\left\langle
b_E \sum_{J=1}^{N_R} y_J
\right\rangle\!\!\!\right\rangle_{\!\! N_R}
-
\left\langle\!\!\!\left\langle
b_E \sum_{I=1}^{N_L} x_I
\right\rangle\!\!\!\right\rangle_{\!\! N_L}
\left\langle\!\!\!\left\langle
\alpha_3 + b_E \sum_{J=1}^{N_R} y_J
\right\rangle\!\!\!\right\rangle_{\!\! N_R}.
\end{split}
\ee
We can reproduce
the exact form of the Nekrasov function \eqref{NekA10} and \eqref{NekA01}
if we decompose \eqref{A1-0} in a rather non-trivial way:
\bel{A10}
\mathcal{A}_{(1),(0)}
= -
\left\{ \alpha_2 + \left( 1 - 
\frac{Q_E}{\alpha_I - Q_E}
\right)
\left\langle\!\!\!\left\langle
b_E \sum_{I=1}^{N_L} x_I 
\right\rangle\!\!\!\right\rangle_{\!\! N_L}
 \right\}
\left\langle\!\!\!\left\langle
b_E \sum_{J=1}^{N_R} y_J  
\right\rangle\!\!\!\right\rangle_{\!\! N_R},
\ee
\bel{A01}
\mathcal{A}_{(0),(1)}
= -
\left\langle\!\!\!\left\langle
b_E \sum_{I=1}^{N_L} x_I
\right\rangle\!\!\!\right\rangle_{\!\! N_L}
\left\{
\alpha_3 +
\left( 1 + \frac{Q_E}{\alpha_I - Q_E} \right)
\left\langle\!\!\!\left\langle
b_E \sum_{J=1}^{N_R} y_J
\right\rangle\!\!\!\right\rangle_{\!\! N_R} \right\}.
\ee
See Appendix \ref{A1detail} for details.

In the next order, the expansion coefficients $\mathcal{A}_2$ can be 
decomposed into five terms
\bel{A2five}
\mathcal{A}_2 = \sum_{|Y_1|+|Y_2|=2} \mathcal{A}_{Y_1,Y_2}=
\mathcal{A}_{(2),(0)} 
+ \mathcal{A}_{(1^2), (0)}
+ \mathcal{A}_{(1),(1)} 
+ \mathcal{A}_{(0),(1^2)} 
+ \mathcal{A}_{(0),(2)},
\ee
each of which can be written as the factorized form 
\bel{PolyMM}
\mathcal{A}_{Y_1, Y_2}
= \Bigl\langle\!\! \Bigl\langle
\ M_{Y_1,Y_2}(x) \ 
\Bigr\rangle\! \! \Bigr\rangle_{\! N_L}
\Bigl\langle\!\! \Bigl\langle
\ \widetilde{M}_{Y_1,Y_2}(y) \ 
\Bigr\rangle\! \! \Bigr\rangle_{\! N_R}
\ee
for some symmetric polynomials $M_{Y_1,Y_2}(x)$ and $\widetilde{M}_{Y_1,Y_2}(y)$.
At this moment, we do not have a 
clear systematic understanding on obtaining these polynomials. But at least for $|Y_1|+|Y_2|\leq 2$,
we obtained explicit forms for them. See Appendix \ref{A2detail}.
Suppose that one of the partition, say $Y_2$, is empty (\,$Y_2=(0)$\, ).
The mass parameters $m_3$ and $m_4$ appear in $\mathcal{A}^{\mathrm{Nek}}_{Y_1,(0)}$ 
through the following factor
\be
\prod_{(i,j) \in Y_1}
(\, a + m_3 + \epsilon_1( i - 1) + \epsilon_2 (j-1)\, )\, 
(\, a + m_4 + \epsilon_1( i - 1) + \epsilon_2 (j-1) ).
\ee
By comparing this with \eqref{GTPL}, it agrees if $\lambda=Y_1$.
Hence we conclude that $\widetilde{M}_{Y_1,(0)}(y)$ is proportional to the Jack polynomial
$P_{Y_1}^{(1/b_E^2)}(y)$. On the other hand, $M_{Y_1,(0)}(x)$ is an 
inhomogeneous symmetric polynomial of $x=(x_1,\dotsc, x_{N_L})$
whose highest degree is equal to $|Y_1|$.
At $b_E=1$ the form of $M_{Y_1,(0)}(x)$ drastically simplifies
and the dominant term is given by the monomial symmetric polynomial:
\be
M_{Y_1,(0)}(x) = c_{Y_1'} \, m_{Y_1'}(x) + \dotsm, \qq
\mbox{at}\ \ b_E=1.
\ee
Here $Y_1'$ is the conjugate partition of $Y_1$
and $c_{Y_1'}$ is a constant.

\section{Some Computations from Loop Equation at Finite $N$ }

In the previous sections, we used the formula \eqref{MK} to
calculate various objects. 
We show that some of results can be obtained
by more standard matrix model technology.
Using the loop equations,
the moments $\langle \!\langle \sum_I x_I \rangle \!\rangle_{N_L}$
and $\langle \!\langle \sum_J y_J \rangle\!\rangle_{N_R}$ can be determined
exactly at finite $N$ without using any approximation.

\subsection{loop equation at finite $N$}

Let us consider the loop equation  at finite $N$ of the perturbed Selberg model at the decoupling limit.
For definiteness, we consider the left-part:
\bel{pSL}
Z_{\mathrm{Selberg}}
(b_E; N_L, \alpha_1, \alpha_2)
= \left( \prod_{I=1}^{N_L} \int_0^1 \de x_I \right)
\prod_{1 \leq I < J \leq N_L} | x_I - x_J|^{2b_E^2}
\exp\left( b_E \sum_{I=1}^{N_L} \widetilde{W}(x_I) \right),
\ee
where
\bel{wtW}
\widetilde{W}(x) := \alpha_1 \log x + \alpha_2 \log (1-x).
\ee
Note that here we do not introduce small $g_s$ parameter which is standard
in the matrix model technology. The standard normalization is $\widetilde{W}(x)
= (1/g_s) W(x)$. But we do not need to consider $g_s$-expansion
or large $N_L$ limit. In the next section, we write it in a more
standard form which is suited for the $g_s$-expansion.

As is usual, by inserting
\be
\sum_{I=1}^{N_L} \frac{\partial}{\partial x_I} \frac{1}{z - x_I}
\ee
into the integrand,  we obtain the loop equation at finite $N$:
\be \label{FiniteLoop}
\left\langle \!\!\! \left\langle \Bigl( \widehat{w}_{N_L}(z) \Bigr)^2
\right\rangle\!\!\!\right\rangle_{\!\! N_L}
+ \left( \widetilde{W}'(z) + Q_E \frac{\de }{\de z} \right)
\left\langle \!\! \left\langle \Bigl. \widehat{w}_{N_L}(z) \Bigr.
\right\rangle\!\! \right\rangle_{\!\! N_L} - \tilde{f}_{N_L}(z) = 0,
\ee
where
\be
\widehat{w}_{N_L}(z):= b_E \sum_{I=1}^{N_L} \frac{1}{z- x_I},
\qq
\tilde{f}_{N_L}(z):=
\left\langle\!\!\!\left\langle b_E \sum_{I=1}^{N_L}
\frac{ \widetilde{W}'(z) - \widetilde{W}'(x_I)}
{z - x_I }
\right\rangle\!\!\!\right\rangle_{\!\! N_L}.
\ee
The loop equation at finite $N$ \eqref{FiniteLoop} is valid for any potential
$\widetilde{W}(z)$ provided that the integral is well-defined.
The expectation value of $\widehat{w}_{N_L}(z)$ is the finite $N$ resolvent:
\bel{FNresol}
\widetilde{w}_{N_L}(z) := \left\langle \!\! \left\langle \Bigl. \widehat{w}_{N_L}(z)
\right\rangle\!\!\right\rangle_{\!\! N_L}
= \left\langle \!\!\! \left\langle b_E
\sum_{I=1}^{N_L} \frac{1}{z - x_I}
\right\rangle\!\!\!\right\rangle_{\!\! N_L} \;.
\ee

For the case of our interest, the potential \eqref{wtW}
takes a very special form:
double-log or two-Penner type. Because  
$\widetilde{W}'(z) = \alpha_1/z + \alpha_2/(z-1)$,
we can see that
\be
\tilde{f}_{N_L}(z) = \frac{c}{z} + \frac{c'}{z-1},
\ee
where the two constants $c$ and $c'$ are given respectively by
\be
c = - \alpha_1
\left\langle\!\!\!\left\langle b_E \sum_{I=1}^{N_L}
\frac{1}{x_I }  
\right\rangle\!\!\!\right\rangle_{\!\! N_L}
= \alpha_1 \widetilde{w}_{N_L}(0), \qq
c' = \alpha_2
\left\langle\!\!\!\left\langle b_E \sum_{I=1}^{N_L}
\frac{ 1}{1 - x_I}
\right\rangle\!\!\!\right\rangle_{\!\! N_L}=
\alpha_2 \widetilde{w}_{N_L}(1).
\ee
Note that
\be \label{moments}
\widehat{w}_{N_L}(z)
= b_E \sum_{I=1}^{N_L} \frac{1}{z - x_I}
= \frac{b_E N_L}{z} + b_E \sum_{\ell=1}^{\infty} \frac{1}{z^{\ell+1}}\,
p_{(\ell)}(x),
\ee
where $p_{(\ell)}(x)$ is the power sum symmetric function of $x_I$:
\be
p_{(\ell)}(x) = \sum_{I=1}^{N_L} x_I^{\ell}.
\ee

By substituting the expansion \eqref{moments} into the loop equation at finite $N$ \eqref{FiniteLoop},
 we obtain infinitely many equations
for the averaging of the symmetric functions of $x_I$.
At $O(1/z)$, we have $c+c'=0$. At $O(1/z^2)$, this constant is
determined exactly as
\be
c' = b_E N_L( \alpha_1 + \alpha_2 + b_E N_L - Q_E).
\ee
At $O(1/z^3)$, we obtain the exact expression for the average of $p_{(1)}(x)$:
\be
\bigl\langle\!\bigl\langle b_E \, p_{(1)}(x)
\bigr\rangle \!\bigr\rangle_{N_L}
= \frac{b_E N_L( \alpha_1 + b_E N_L - Q_E )}{\alpha_1 + \alpha_2 + 2 b_E N_L - 2 Q_E },
\ee
which is consistent with \eqref{mmt1}.
At $O(1/z^4)$ the loop equation \eqref{FiniteLoop} at finite $N$ contains two
unknown moments $\langle \! \langle \, b_E \, p_{(2)} \, \rangle\!\rangle_{N_L}$
and $\langle \! \langle (b_E \, p_{(1)})^2 \rangle\!\rangle_{N_L}$:
\be
\begin{split}
& ( \alpha_1 + \alpha_2 + 2 b_E N_L - 3 Q_E) \,
\bigl\langle\!\bigl\langle b_E \, p_{(2)}(x)
\bigr\rangle \!\bigr\rangle_{N_L}
+ \bigl\langle\!\bigl\langle \, \bigl( b_E\, p_{(1)}(x) \bigr)^2
\bigr\rangle \!\bigr\rangle_{N_L} \cr
&= b_E N_L ( \alpha_1 + b_E N_L - Q_E)
- \alpha_2
\bigl\langle\!\bigl\langle b_E \, p_{(1)}(x)
\bigr\rangle \!\bigr\rangle_{N_L}.
\end{split}
\ee
Hence beyond $O(1/z^4)$, the loop equation
alone is not sufficient to get
exact non-perturbative results. But it is sufficient for obtaining
$\mathcal{B}_1$ and $\mathcal{A}_1$.

 We summarize the exact results obtained from 
the loop equation at finite $N$ \eqref{FiniteLoop} alone:
\bel{NPRL}
\begin{split}
\Bigl\langle\!\!\Bigl\langle b_E \, p_{(1)}(x)
\Bigr\rangle \!\!\Bigr\rangle_{\! N_L}
=
\left\langle\!\!\!\left\langle
b_E \sum_{I=1}^{N_L} x_I
\right\rangle\!\!\!\right\rangle_{\!\! N_L}
&= \frac{b_E N_L( b_E N_L - Q_E + \alpha_1)}{(\alpha_1 + \alpha_2 + 2 b_E N_L - 2 Q_E)}, \cr
\tilde{f}_{N_L}(z) &=  \frac{b_E N_L(\alpha_1+\alpha_2 + b_E N_L - Q_E)}{z(z-1)}, \cr
- \widetilde{w}_{N_L}(0)=
\left\langle\!\!\!\left\langle
b_E \sum_{I=1}^{N_L} \frac{1}{x_I}
\right\rangle\!\!\!\right\rangle_{\!\! N_L}
&=\frac{b_E N_L(\alpha_1+\alpha_2 + b_E N_L - Q_E)}{\alpha_1}, \cr
\widetilde{w}_{N_L}(1)=
\left\langle\!\!\!\left\langle
b_E \sum_{I=1}^{N_L} \frac{1}{1 - x_I}
\right\rangle\!\!\!\right\rangle_{\!\! N_L}
&=\frac{b_E N_L(\alpha_1+\alpha_2 + b_E N_L - Q_E)}{\alpha_2}.
\end{split}
\ee
Except the first equation, which is \eqref{mmt1},
the remainder is a set of independent non-perturbative
results for the perturbed Selberg model.

\section{Computation at the Planar Level and $g_s$ Corrections}

In this section, we consider  the $g_s$-expansion of the loop equation \eqref{FiniteLoop}.
For this purpose, we introduce the scale $g_s$ and rescale various parameters.
First, we set
\be
\tilde{\alpha}_i:= g_s \alpha_i, \qq i=1,2,3,4.
\ee
Then, the potential is rewritten as
\bel{MMpot}
\widetilde{W}(x) \equiv \frac{1}{g_s} W(x),
\qq
W(x) := \tilde{\alpha}_1 \log x + \tilde{\alpha}_2 \log(1-x).
\ee
Now the perturbed Selberg model takes the standard form in matrix model
calculation:
\be
Z_L = \left( \prod_{I=1}^{N_L} \int_0^1 \de x_I \right)
\, \prod_{1 \leq I < J \leq N_L} | x_I - x_J |^{2b_E^2}
\exp\left( \frac{b_E}{g_s} \sum_{I=1}^{N_L} W(x_I) \right).
\ee
A similar expression holds for the right part.

We will use the 't Hooft coupling $S_L$ and $S_R$ and their deformed version:
\bel{tHooft}
S_L:= g_s N_L, \qq
S_R:= g_s N_R, \qq
\widetilde{S}_L:= g_s b_E N_L, \qq
\widetilde{S}_R:= g_s b_E N_R.
\ee
In the rescaled parameters, the momentum conservation condition \eqref{MCC2}
turns into
\bel{RMCC}
\tilde{\alpha}_1 + \tilde{\alpha}_2 + \tilde{\alpha}_3 + \tilde{\alpha}_4
+ 2 \widetilde{S}_L + 2 \widetilde{S}_R = 2 \epsilon.
\ee

We also change the normalization of the resolvent
$w_{N_L}(z):= g_s \widetilde{w}_{N_L}(z)$,
\bel{resolFN}
w_{N_L}(z)=
\left\langle\!\!\!\left\langle b_E g_s \sum_{I=1}^{N_L}
\frac{1}{z-x_I}
\right\rangle\!\!\!\right\rangle_{N_L}, \qq
w_{N_R}(z):=
\left\langle\!\!\!\left\langle b_E g_s \sum_{I=1}^{N_R}
\frac{1}{z-y_I}
\right\rangle\!\!\!\right\rangle_{N_R}.
\ee
Also, we set
\bel{fNL2}
f_{N_L}(x):= g_s^2 \tilde{f}_{N_L}(x)
=  
\left\langle\!\!\!\left\langle b_E \, g_s \sum_{I=1}^{N_L}
\frac{ W'(z) - W'(x_I)}
{z - x_I }
\right\rangle\!\!\!\right\rangle_{\!\! N_L}.
\ee
Now the loop equation at finite $N$ takes a more standard form:
\bel{FL2}
\left\langle\!\!\!\left\langle
\left( b_E g_s \sum_{I=1}^{N_L} \frac{1}{z-x_I} \right)^2
\right\rangle\!\!\!\right\rangle_{\!\!N_L}
+
W'(z) w_{N_L}(z)+ \epsilon \,  
w'_{N_L}(z) - f_{N_L}(z) = 0.
\ee
Here $\epsilon = g_s Q_E = \epsilon_1 + \epsilon_2$.
A similar expression holds for the right part.

\subsection{$b_{E}=1$ $(\epsilon=0)$}

We first consider $b_E=1$ case.
In the planar limit, \eqref{FL2} reduces to an algebraic equation
\be \label{Eq11}
w_L(z)^2 + W'(z) \, w_L(z) - f(z) = 0,
\ee
where
\be
w_L(z) =
\lim_{\stackrel{\scr N_L \rightarrow \infty}{g_s \rightarrow 0}}
w_{N_L}(z), \qq
f(z) =
\lim_{\stackrel{\scr N_L \rightarrow \infty}{g_s \rightarrow 0}}
f_{N_L}(z).
\ee
In the planar limit, the momentum conservation condition \eqref{RMCC}
becomes
\be
\tilde{\alpha}_1 + \tilde{\alpha}_2 + \tilde{\alpha}_3 + \tilde{\alpha}_4
+ 2 S_L + 2 S_R = 0.
\ee

The solution to the planar loop equation \eqref{Eq11} is
\be \label{Eq4-1}
w_L(z) = - \frac{W'(z)}{2} + \sqrt{ \left( \frac{W'(z)}{2} \right)^2 + f(z)}.
\ee
The form of $f(z)$ is already restricted to $\frac{c}{z-1} + \frac{c'}{z}$ and
the residue at infinity, which is proportional to $c+c'$, vanishes due to
the stationary condition on the distribution of the eigenvalues. Hence
\be \label{Eq4-2}
f(z) = \frac{c}{z(z-1)}.
\ee
Picking the residue of $w_L(z)$ at infinity, and equating it with
$S_L=g_s N_L$,
we obtain
\be \label{Eq4-3}
c = S_L^2 + ( \tilde{\alpha}_1 + \tilde{\alpha}_2 ) S_L.
\ee
The solution takes the form of
\be \label{Eq4-4}
w_L(z) = w(z ; \tilde{\alpha}_1, \tilde{\alpha}_2, S_L)
= - \frac{1}{2} \left( \frac{\tilde{\alpha}_1}{z}
+ \frac{\tilde{\alpha_2}}{z-1} \right)
+ \sqrt{ \frac{1}{4} \left( \frac{\tilde{\alpha}_1}{z}
+ \frac{\tilde{\alpha}_2}{z-1} \right)^2
+ \frac{S_L(\tilde{\alpha}_1 + \tilde{\alpha}_2+ S_L)}
{ z(z-1)} }
\ee
and is completely determined. Likewise,
\be \label{Eq4-5}
w_R(z) = w(z; \tilde{\alpha}_4, \tilde{\alpha}_3, S_R)
= - \frac{1}{2} \left(
\frac{\tilde{\alpha}_4}
{z} + \frac{\tilde{\alpha}_3}{z-1} \right)
+ \sqrt{ \frac{1}{4} \left( \frac{\tilde{\alpha}_4}{z}
+ \frac{\tilde{\alpha}_3}{z-1} \right)^2
+ \frac{S_R( \tilde{\alpha}_4+ \tilde{\alpha}_3 + S_R)}{z(z-1)}}.
\ee
We expand them as follows
\bel{wLRl}
w_L(z) = \sum_{\ell=0}^{\infty} \frac{w_L^{(\ell)}}{z^{\ell+1}}, \qq
w_R(z) = \sum_{\ell=0}^{\infty} \frac{w_R^{(\ell)}}{z^{\ell+1}}.
\ee
They correspond to
\bel{plam}
w_L^{(\ell)} = \lim_{\stackrel{\scr N_L \rightarrow \infty}{g_s \rightarrow 0}}
\left\langle\!\!\!\left\langle
g_s \sum_{I=1}^{N_L} x_I^{\ell}
\right\rangle\!\!\!\right\rangle_{N_L},
\qq
w_R^{(\ell)} = \lim_{\stackrel{\scr N_R \rightarrow \infty}{g_s \rightarrow 0}}
\left\langle\!\!\!\left\langle
g_s \sum_{I=1}^{N_R} y_I^{\ell}
\right\rangle\!\!\!\right\rangle_{N_R}.
\ee

At the first order, the coefficient of \eqref{PleA} and  that of \eqref{PleB}
are respectively
\be
\mathcal{A}_1 =
- \frac{1}{g_s^2} ( w_L^{(1)} + \tilde{\alpha}_2)
w_R^{(1)} - \frac{1}{g_s^2} w_L^{(1)} ( w_R^{(1)} + \tilde{\alpha}_3),
\ee
\be
\mathcal{B}_1 = -2 \left( w_L^{(1)} + \frac{1}{2} \tilde{\alpha}_2 \right)
\left( w_R^{(1)} + \frac{1}{2} \tilde{\alpha}_3 \right)
= \mathcal{A}_1 - \frac{1}{2 g_s^2} \tilde{\alpha}_2 \tilde{\alpha}_3.
\ee
For simplicity, we use
\be
\tilde{\alpha}_I:= \tilde{\alpha}_1 + \tilde{\alpha}_2 + 2 S_L
= - \tilde{\alpha}_4 - \tilde{\alpha}_3 - 2 S_L.
\ee
At the first order, we obtain
\bel{pmmt1}
w_L^{(1)} = \frac{S_L(\tilde{\alpha}_1 + S_L)}
{\tilde{\alpha}_I},
\qq
w_R^{(1)} = - \frac{S_R(\tilde{\alpha}_4 + S_R)}
{\tilde{\alpha}_I}.
\ee
Thus
\bel{pmmt1b}
w_L^{(1)} +  \tilde{\alpha}_2
= \frac{(\tilde{\alpha}_2 + S_L)(\tilde{\alpha}_I - S_L)}
{\tilde{\alpha}_I},
\qq
w_R^{(1)} +  \tilde{\alpha}_3 =
 \frac{( \tilde{\alpha}_3 + S_R)( \tilde{\alpha}_I + S_R)}
{ \tilde{\alpha}_I},
\ee
and
\be \label{Eq4-8}
\mathcal{A}_1 =
\frac{S_R( \tilde{\alpha}_4 + S_R)
(\tilde{\alpha}_2 + S_L) ( \tilde{\alpha}_I - S_L) }
{g_s^2 \, \tilde{\alpha}_I^2}
- \frac{S_L(\tilde{\alpha}_1 + S_L)( \tilde{\alpha}_3 + S_R)
(\tilde{\alpha}_I + S_R)}{g_s^2 \, \tilde{\alpha}_I^2}.
\ee
Adding $-(1/2)\tilde{\alpha}_2 \tilde{\alpha}_3/g_s^2$ coming from the
Veneziano factor $q^{(1/2)\alpha_1 \alpha_2} (1-q)^{(1/2) \alpha_2 \alpha_3}$,
we obtain
\be \label{Eq4-9}
\mathcal{B}_1 =
\frac{1}{g_s^2\, \tilde{\alpha}_I^2}
\Bigl( 2 S_L (\tilde{\alpha}_1 + \tilde{\alpha}_2 + S_L) + \tilde{\alpha}_2
( \tilde{\alpha}_1 + \tilde{\alpha}_2 ) \Bigr)\,  
\Bigl( 2 S_R ( \tilde{\alpha}_4 + \tilde{\alpha}_3 + S_R)
+ \tilde{\alpha}_3( \tilde{\alpha}_4 + \tilde{\alpha}_3) \Bigr).
\ee

In the planar limit, the higher coefficients $\mathcal{A}_{\ell}$
and $\mathcal{B}_{\ell}$ ($\ell \geq 2$) are factorized as follows:
\be
\mathcal{A}_{\ell} = \frac{1}{\ell!} ( \mathcal{A}_1)^{\ell}, \qq
\mathcal{B}_{\ell} = \frac{1}{\ell!} ( \mathcal{B}_1)^{\ell}.
\ee
The planar contributions to $\mathcal{A}_{\ell}$ and $\mathcal{B}_{\ell}$
exponentiate to give
\be
\mathcal{A}_{\mathrm{planar}}(q)
= \sum_{\ell=0}^{\infty} q^{\ell}  \,  \mathcal{A}_{\ell}
= \exp\left(q \mathcal{A}_1\right), \qq
\mathcal{B}_{\mathrm{planar}}(q)
= \sum_{\ell=0}^{\infty} q^{\ell} \, \mathcal{B}_{\ell}
= \exp\left( q \mathcal{B}_1 \right).
\ee
The higher moments \eqref{plam} in the planar limit
are, therefore, not really relevant for the calculation of
$\mathcal{A}_{\ell}$ and $\mathcal{B}_{\ell}$ $(\ell \geq 2)$,
but are useful for the calculation of the planar free energy.
For simplicity, let us consider
\be
\mathcal{A}(q)
= \exp\left( \mathcal{F}(q) \right)
= \exp\left( \frac{1}{g_s^2} \mathcal{F}_0(q)
+ \mathcal{F}_1(q) + g_s^2 \mathcal{F}_2(q) + \dotsm \right).
\ee
Then, we can easily evaluate
the $q$-expansion coefficients of the planar free energy $\mathcal{F}_0(q)$
\be
\mathcal{F}_0(q) = \sum_{k=1}^{\infty} q^{k}
\mathcal{F}_{0,k}.
\ee
We already have
\be
\mathcal{F}_{0,1} = g_s^2 \mathcal{A}_1 =
\frac{S_L(\tilde{\alpha}_1 + S_L)( \tilde{\alpha}_3 + S_R)
(\tilde{\alpha}_I - S_R)}{\tilde{\alpha}_I^2}
- \frac{(S_L + \tilde{\alpha}_2) ( \tilde{\alpha}_I - S_L) S_R
( \tilde{\alpha}_4 + S_R)}{\tilde{\alpha}_I^2}.
\ee
From \eqref{PleA}, we easily see that
\bel{F0k} 
\mathcal{F}_{0, k}
= - \frac{1}{k}
\left[
( w_L^{(k)} + \tilde{\alpha}_2 ) w_R^{(k)}
+ w_L^{(k)} ( w_R^{(k)} + \tilde{\alpha}_3)
\right].
\ee

In the next order
\be \label{Eq4-10}
w_L^{(2)}
= \frac{ S_L (\tilde{\alpha}_1 + S_L )
\{
\tilde{\alpha}_1( \tilde{\alpha}_1 + \tilde{\alpha}_2)
+ S_L( 3 \tilde{\alpha}_1 + 2 \tilde{\alpha}_2 + 3 S_L) \}}
{ \tilde{\alpha}_I^3},
\ee
and  a similar expression for $w_R^{(2)}$  is obtained by
 the replacements
\be
S_L \longrightarrow S_R, \qq
\tilde{\alpha}_1 \longrightarrow \tilde{\alpha}_4, \qq
\tilde{\alpha}_2 \longrightarrow \tilde{\alpha}_3, \qq
\tilde{\alpha}_I \longrightarrow - \tilde{\alpha}_I.
\ee
Then
\bel{pmmt2b}
w_L^{(2)} + \tilde{\alpha}_2
= \frac{(\tilde{\alpha}_2 + S_L)( \tilde{\alpha}_I - S_L)
[ \tilde{\alpha}_I( \tilde{\alpha}_I - S_L) + S_L( \tilde{\alpha}_2 + S_L)]}
{\tilde{\alpha}_I^3},
\ee
and a similar expression  is obtained for $w_R^{(2)} + \tilde{\alpha}_3$.
By substituting these relations into
\be
\mathcal{F}_{0,2} = - \frac{1}{2} \Bigl[
( w_L^{(2)} + \tilde{\alpha}_2) w_R^{(2)}
+ w_L^{(2)} ( w_R^{(2)} + \tilde{\alpha}_3) \Bigr],
\ee
we can obtain an explicit form of $\mathcal{F}_{0,2}$.

In this way, $w_L^{(k)}$ and $w_R^{(k)}$ for $k=1,2,3,\dotsc$
can be computed and upon substitution into \eqref{F0k},
$\mathcal{F}_{0,k}$ are determined.
The $q$-expansion coefficients of the planar free energy
are completely controlled by
the single algebraic function $w(z; \alpha, \beta, S)$.

We can match parameters and can get planar 0d-4d relations
also by using the planar loop equations.
For example, let us  match parameters of matrix model with those of CFT.
In the planar limit, we have an expression of $\mathcal{B}_1$ \eqref{Eq4-9}.
On the other hand,
\bel{CFTB1b}
\mathcal{B}_1^{(\mathrm{CFT})} =
\frac{(\Delta_I + \Delta_2 - \Delta_1)(\Delta_I + \Delta_3 - \Delta_4)}
{2 \Delta_I}.
\ee
We know that in the $c=1$ free boson system, the vertex operator
$: \ex^{(\tilde{\alpha}_i/2 g_s) \phi(z)} :$
has the scaling dimension $\Delta_i = 1/(4g_s^2) \tilde{\alpha}_i^2$.
By identifying the internal momenta with $\alpha_I =(1/g_s)\tilde{\alpha}_I$,
\eqref{Eq4-9} completely agrees with the CFT expression \eqref{CFTB1b}
\footnote{If $g_s \rightarrow 0$ while keeping $\tilde{\alpha}_i$  finite,
the external vertex operators carry very large momenta $\alpha_i = (1/g_s) \tilde{\alpha}_i$.}.
 As for the gauge theory parameters, analysis can be done in the same way as that of section
\ref{ChapMMvsNek}. The planar 0d-4d relation is
\begin{align} \label{P0d4d}
S_L &= a - m_2,& 
S_R &= - (a + m_3), \cr
\tilde{\alpha}_1 &= m_2 - m_1,&
\tilde{\alpha}_2 &= m_2 + m_1, \cr
\tilde{\alpha}_3 &= m_3 + m_4,&
\tilde{\alpha}_4 &= m_3 - m_4.
\end{align}
Also, $\tilde{\alpha}_I = 2a$.
Using the relation \eqref{P0d4d}, the solution \eqref{Eq4-4}
to the planar loop equation takes the form
\bel{wLx2}
w_L(x) = - \frac{1}{2} \left( \frac{m_2 - m_1}{z}
+ \frac{m_2 + m_1}{z-1} \right)
+ \sqrt{\frac{1}{4}
\left( \frac{m_2 - m_1}{z}
+ \frac{m_2 + m_1}{z-1} \right)^2
+ \frac{(a+m_2)(a-m_2)}{z(z-1)}}.
\ee
A similar expression for $w_R(x)$ can be obtained by the replacement
\eqref{GTparity}.

First few moments \eqref{pmmt1}, \eqref{pmmt1b}, \eqref{Eq4-10}
and \eqref{pmmt2b}
are rewritten as follows
\be
\begin{split}
w_L^{(1)} &= \frac{(a-m_1)(a-m_2)}{2a},  \cr
w_L^{(1)} + \tilde{\alpha}_2 &= \frac{(a+m_1)(a+m_2)}{2a}, \cr
w_L^{(2)} &= \frac{(a-m_1)(a-m_2)}{2a} - \frac{(a+m_1)(a+m_2)(a-m_1)(a-m_2)}{8a^3}, \cr
w_L^{(2)} + \tilde{\alpha}_2
&= \frac{(a+m_1)(a+m_2)}{2a} - \frac{(a+m_1)(a+m_2)(a-m_1)(a-m_2)}{8a^3},
\end{split}
\ee 
and similar expressions for the right part.

First two of the planar instanton contribution \eqref{F0k}
obtained from the expansion coefficients of \eqref{wLx2} are
\be
\mathcal{F}_{0,1} = 
\frac{(a+m_1)(a+m_2)(a+m_3)(a+m_4)}{4a^2} 
+ \frac{(a-m_1)(a-m_2)(a-m_3)(a-m_4)}{4a^2},
\ee
\be
\begin{split}
\mathcal{F}_{0,2} 
&= \frac{\prod_{i} ( a + m_i)}{8a^2}
\left( 1 - \frac{(a-m_1)(a-m_2)}{4a^2} \right)
\left( 1 - \frac{(a-m_3)(a-m_4)}{4a^2} \right) \cr
& + \frac{\prod_{i} ( a - m_i)}{8a^2}
\left( 1 - \frac{(a+m_1)(a+m_2)}{4a^2} \right)
\left( 1 - \frac{(a+m_3)(a+m_4)}{4a^2} \right).
\end{split}
\ee


\subsection{$b_{E} \neq 1$: $g_s$-corrected resolvent}

We briefly comment on the case of $b_E \neq 1$
and on the $g_s$-correction\footnote{
It is known that \cite{har}
when $b_E \neq 1$, the $g_s$-expansion contains odd powers in $g_s$.}
to the resolvent \eqref{resolFN}.
We choose $g_s$ and $b_E$
as independent parameters
and consider the perturbation of the loop equation \eqref{FL2} in $g_s$. 
The planar results for $b_E \neq 1$ can be obtained from
those of $b_E=1$ by replacing the 't Hooft parameters $S_L$, $S_R$
by the deformed 't Hooft parameters 
$\widetilde{S}_L$, $\widetilde{S}_R$ respectively \cite{IMO}.
(See \eqref{tHooft}).

In the $g_s$-expansion, some care is necessary
on the choice of parameters to be fixed. We would like to compare
the $g_s$-expansion coefficients with those of the gauge theory.
We have an exact 0d-4d relation \eqref{0d4d} and we fix the
parameters of $\mathcal{N}=2$ gauge theory. 
There is a subtle point. Notice that for $b_E \neq 1$ (i.e., $Q_E \neq 0$), 
the parameters 
in the matrix model potential \eqref{MMpot} are rewritten as
\begin{align}
\tilde{\alpha}_1 &= m_2 - m_1 + g_s Q_E,&
\tilde{\alpha}_2 &= m_2 + m_1, \cr
\tilde{\alpha}_3 &= m_3 + m_4,& 
\tilde{\alpha}_4 &= m_3 - m_4 + g_s Q_E.
\end{align}
The parameters $\tilde{\alpha}_1$ and $\tilde{\alpha}_4$
are inhomogeneous with respect to the degree in $g_s$.
The derivative of the potential has the following $g_s$-expansion
\bel{gsdW}
W'(z) = W_0'(z) + \frac{g_s Q_E}{z}, \qq
W_0'(z) := \frac{m_2-m_1}{z} + \frac{m_2+m_1}{z-1}.
\ee
For simplicity, we consider the left part only.

Recall that we have the exact form of the function $f_{N_L}(z)$ \eqref{fNL2} 
(see \eqref{NPRL}). We see that it receives no $g_s$-correction:
\bel{gsf}
f_{N_L}(z) = \frac{(a+m_2)(a-m_2)}{z(z-1)} = f_{L}(z).
\ee

The resolvent \eqref{resolFN} admits the following $g_s$-expansion
\bel{gsw}
w_{N_L}(z) = w_L(z) + g_s w_L^{(1/2)}(z) + g_s^2 w_L^{(1)}(z) + \dotsm.
\ee
Up to $O(g_s)$, the first term in the loop equation
\eqref{FL2} behaves as
\bel{gsWW}
\left\langle \!\! \left\langle
\bigl( \widehat{w}_{N_L}(z) \bigr)^2
\right\rangle \!\! \right\rangle_{N_L}
= w_{L}(z)^2 + O(g_s^2).
\ee
By substituting \eqref{gsdW}, \eqref{gsf}, \eqref{gsw} and \eqref{gsWW} into
the loop equation \eqref{FL2}, we can easily see that the $O(g_s)$ relation is
given by 
\be
\bigl( 2 w_L(z) + W_0'(z) \bigr) w_L^{(1/2)}(z) 
+ \frac{Q_E}{z} w_L(z) + Q_E \, w_L'(z) = 0.
\ee
Hence the $O(g_s)$ corrected part is easily determined as
\be
w_L^{(1/2)}(z)
= - \frac{Q_E}{z( 2 w_L(z) + W_0'(z))} \frac{\de}{\de z} \bigl( z \, w_L(z) \bigr).
\ee
Therefore, the $g_s$-correction is also characterized by
the algebraic function $w(z; \alpha, \beta, S)$. Here
$w_L(z)$ takes exactly the same form as the planar solution 
\eqref{wLx2} to $b_E=1$ case.


\section{Free Field Representation of Nekrasov Function}

Finally, we come back to the definition of the perturbed double-Selberg model \eqref{Z3P0}
given in terms of the free chiral boson.
Using this definition,  we can write a free field representation
of the Nekrasov function.
We write the modes of the free boson as
\be
\phi(z) = \phi_0 + a_0 \log z - \sum_{n \neq 0} \frac{a_n}{n} z^{-n}
= \phi_0 + a_0 \log z + \phi_+(z) + \phi_-(z),
\ee
\be
\phi_+(z) = - \sum_{n > 0} \frac{a_n}{n} z^{-n}, \qq
\phi_-(z) = \sum_{n>0} \frac{a_{-n}}{n} z^n.
\ee
The non-trivial commutation relations are (cf. footnote 1)
\be
[ a_0, \phi_0 ] = 2, \qq
[a_n, a_m ] = 2 \,n \, \delta_{n+m,0}.
\ee

We decompose the Virasoro generator
$L_0$ 
into the non-zero mode part and  the zero-mode part:
\be
L_0 = \widehat{K} + \widehat{\Delta},
\qq
\widehat{K} := \frac{1}{2} \sum_{n \geq 1} a_{-n}\, a_n, \qq
\widehat{\Delta}:= \frac{1}{4} a_0^2 - \frac{Q_E}{2}\, a_0.
\ee 
The eigenstate of $a_0$ is defined by
\be
| \alpha \rangle:= \ex^{(\alpha/2) \phi_0} \, | \,0\, \rangle, \qq
\langle \alpha | := \langle \,0\, | \, \ex^{-(\alpha/2) \phi_0}.
\ee
Let us introduce ``left'' intermediate state $|\alpha_I; q \rangle$
and ``right'' intermediate state $\langle \alpha_I |$ by
\be
|\, \alpha_I; q \rangle \equiv
: \ex^{(1/2) \alpha_2 \phi(q)}:
(Q_{[0,q]})^{N_L}\, 
\bigl| \, \alpha_1 \bigr\rangle,
\qq
\langle \alpha_I | \equiv 
\langle 2 Q_E - \alpha_4 \, |\, ( Q_{[1,\infty]})^{N_R}
\, : \ex^{(1/2)\alpha_3 \phi(1)}:, 
\ee
where
\be
Q_{[z_1,z_2]}= \int_{z_1}^{z_2} \de z : \ex^{b_E \phi(z)} :.
\ee 
Note that
\be
a_0 | \, \alpha_I; q \rangle = | \, \alpha_I; q \rangle \, \alpha_I, \qq
\alpha_I = \alpha_1 + \alpha_2 + 2 b_E N_L,
\ee
\be
\widehat{\Delta} | \, \alpha_I; q \rangle
= | \, \alpha_I; q \rangle \, \Delta_I, \qq
\Delta_I = \frac{1}{4}\alpha_I(\alpha_I-2Q_E).
\ee 
The definition of the perturbed double-Selberg/three-Penner
matrix model \eqref{Z3P0} can be concisely written as
\be
Z_{\mathrm{pert}-(\mathrm{Selberg})^2} 
= \langle \alpha_I \, | \, \alpha_I; q \, \rangle.
\ee
The dependence on the cross ratio $q$ appears only 
in the left intermediate state.

Recall that the vertex operator transforms under the scale transformation
\be
q^{L_0} \bigl( : \ex^{(\alpha/2) \phi(z) }:  \bigr) q^{-L_0}
= q^{\Delta_{\alpha}} : \ex^{(\alpha/2) \phi(qz) }:.
\ee
The ``left'' screening charge $Q_{[0,q]}$
behaves under the coordinate transformation
$z = q x$ as
\be
Q_{[0,q]}
= q^{L_0}
\left( \int_0^1 \de x : \ex^{b_E \phi(x)}: \right) q^{-L_0}
= q^{L_0} \, Q_{[0,1]} \, q^{-L_0}.
\ee
Hence the left intermediate state can be rewritten as
\be
\begin{split}
| \alpha_I \rangle
&= q^{L_0} \, : \ex^{(1/2) \alpha_2 \phi(1)}:\,  
( Q_{[0,1]})^{N_L}\, 
| \, \alpha_1 \rangle \  q^{-\Delta_1 - \Delta_2} \cr
&= q^{L_0 - \Delta_1 - \Delta_2}
| N_L, \alpha_1, \alpha_2 ; b_E \rangle,
\end{split}
\ee
where
\be
\begin{split}
| \, N_L; \alpha_1, \alpha_2; b_E \, \rangle
&\equiv : \ex^{(1/2) \alpha_2 \phi(1)}:\,  
( Q_{[0,1]})^{N_L}\, 
| \, \alpha_1 \rangle \  \cr
&= \left( \prod_{I=1}^{N_L}
\int_0^1 \de x_I \right) \prod_{I=1}^{N_L} 
x_I^{b_E \alpha_1} ( 1 - x_I)^{b_E \alpha_2}
\prod_{1 \leq I < J \leq N_L} |x_I - x_J |^{2b_E^2} \cr
& \qq \qq \qq \times
\exp\left( \frac{1}{2} \alpha_2 \phi_-(1)
+ b_E \sum_{I=1}^{N_L} \phi_-(x_I) \right)
| \, \alpha_I \rangle.
\end{split}
\ee
We call $|N_L, \alpha_1, \alpha_2 ; b_E \rangle$ Selberg state.
Similarly we can introduce the ``right'' Selberg state
$\langle \, N_R, \alpha_4, \alpha_3; b_E |$.

Then the partition function of the perturbed double-Selberg model
can be compactly written as
\be
Z_{\mathrm{pert}-(\mathrm{Selberg})^2}(q)
= q^{\Delta_I - \Delta_1 - \Delta_2}
\langle N_R, \alpha_4, \alpha_3; b_E |
\, q^{\widehat{K}}\,
| N_L, \alpha_1, \alpha_2; b_E \rangle.
\ee

Furthermore, if we convert matrix model parameters into the those of gauge theory
by the $0d-4d$ relations, we also rewrite it as follows:
\be
\begin{split}
&Z_{\mathrm{pert}-(\mathrm{Selberg})^2} \cr
&= q^{\sigma}\, 
\langle \, 0 \, |\, 
\ex^{(1/2 g_s)( m_3 - m_4 - \epsilon) \phi_0}
\, ( Q_{[1,\infty]})^{- (1/b_E g_s)(a+m_3)}\, 
: \ex^{(1/2g_s) (m_3+m_4) \phi(1)}: \ \cr
& \qq \times 
\, q^{\widehat{K}} \, 
: \ex^{(1/2g_s) (m_2+m_1) \phi(1) }: \, 
( Q_{[0,1]} )^{(1/b_E g_s) ( a - m_2)}\, 
\ex^{(1/2 g_s) ( m_2 - m_1 + \epsilon) \phi_0 } \, | \, 0 \, \rangle.
\end{split}
\ee
This gives a $\beta$-deformed version of the formula  seen in \cite{LMN}.

\section*{Acknowledgements}
We are grateful to Hiroaki Kanno for several insightful remarks and
discussions on the Nekrasov function and Yutaka Matsuo for providing us with some references.
We also thank Andrei Marshakov for interesting discussions and comments on
quiver (conformal) matrix models and Alyosha Morozov and the other participants of
the conference ``Recent Advances in Gauge Theories and CFTs" 
at Yukawa Institute for Theoretical Physics (YITP) which took place
 in the  last stage of the current manuscript.
The research of H.~I.~ and T.~O.~
is supported in part by the Grant-in-Aid for Scientific Research (2054278)
from the Ministry of Education, Science and Culture, Japan.

\appendix

\section{Matrix Model and Nekrasov Partition Function}

\subsection{0d-4d Dictionary}

In this appendix, we summarize the 0d-4d dictionary for the
$\beta$-deformed matrix model (the perturbed double-Selberg model) \eqref{Z3P2}.
The following is our notation for
parameters of 0d matrix model and 4d gauge theory:
\be \label{04C}
Z^{(\mathrm{0d})}_{\mathrm{pert}-(\mathrm{Selberg})^2}
( q\, |\, b_E; N_L, \alpha_1, \alpha_2; N_R, \alpha_4,\alpha_3)
= \mathcal{A}^{(4d)}_{\mathrm{Nek}}
\left(q \left|  \, \frac{\epsilon_1}{g_s};
\frac{a}{g_s}, \frac{m_1}{g_s}, \frac{m_2}{g_s},
\frac{m_3}{g_s}, \frac{m_4}{g_s} \right. \right).
\ee
In addition, on 4d side, the following symbols are often used:
\be
\frac{\epsilon_1}{g_s} \frac{\epsilon_2}{g_s} = - 1, \qq
\epsilon = \epsilon_1 + \epsilon_2.
\ee
The matrix model parameters obey the momentum conservation condition:
\be
\alpha_1 + \alpha_2 + \alpha_3 + \alpha_4 + 2 (N_L + N_R) b_E = 2 Q_E.
\ee
Also, on 0d side, we use
\be
Q_E = b_E - \frac{1}{b_E}, \qq
\alpha_I = \alpha_1 + \alpha_2 + 2 b_E N_L = - \alpha_3 - \alpha_4
- 2 b_E N_R + 2 Q_E.
\ee
The 0d-4d relation is given by
\begin{align} \label{AGT0d4dR}
b_E N_L &= \frac{a-m_2}{g_s},&
b_E N_R &= - \frac{a+m_3}{g_s}, \cr 
\alpha_1 &= \frac{1}{g_s}( m_2 - m_1 + \epsilon ),&
\alpha_2 &= \frac{1}{g_s}( m_2 + m_1 ), \cr
\alpha_3 &= \frac{1}{g_s}( m_3
 + m_4 ),&
\alpha_4 &= \frac{1}{g_s}( m_3 - m_4 + \epsilon ).
\end{align}

\subsection{some relations among parameters}

Some relations among parameters are collected below:
\be
\frac{\epsilon_1}{g_s} = b_E, \qq
\frac{\epsilon_2}{g_s} = - \frac{1}{b_E}, \qq
\frac{\epsilon}{g_s} = \frac{\epsilon_1 + \epsilon_2}{g_s}= Q_E
= b_E - \frac{1}{b_E},
\ee
\be
\frac{a}{g_s} = \frac{1}{2} ( \alpha_I - Q_E) = \frac{1}{2} ( \alpha_1 + \alpha_2
+ 2 b_E N_L - Q_E) = \frac{1}{2} ( - \alpha_3 - \alpha_4 - 2 b_E N_R + Q_E ),
\ee
\begin{align}
\frac{m_1}{g_s}
&= \frac{1}{2} ( \alpha_2 - \alpha_1 + Q_E),&
\frac{m_2}{g_s}
&= \frac{1}{2} ( \alpha_2 + \alpha_1 - Q_E), \cr
\frac{m_3}{g_s}
&= \frac{1}{2} ( \alpha_3+  \alpha_4 - Q_E),&
\frac{m_4}{g_s}
&= \frac{1}{2} ( \alpha_3 - \alpha_4 + Q_E).
\end{align}
\begin{align}
\frac{a+m_1}{g_s} &= \alpha_2 + b_E N_L,&
\frac{a-m_1}{g_s} &= \alpha_1 + b_E N_L - Q_E, \cr
\frac{a+m_2}{g_s} &= \alpha_1 + \alpha_2 + b_E N_L - Q_E,&
\frac{a-m_2}{g_s} &= b_E N_L, \cr
\frac{a+m_3}{g_s} &= - b_E N_R,&
\frac{a-m_3}{g_s} &=-\alpha_3 - \alpha_4 - b_E N_R + Q_E,\cr
\frac{a+m_4}{g_s} &= - \alpha_4 - b_E N_R + Q_E,&
\frac{a-m_4}{g_s} &= - \alpha_3 - b_E N_R.
\end{align}
\be
\begin{split}
\frac{2a+\epsilon}{g_s} &= \alpha_I = \alpha_1 + \alpha_2 + 2 b_E N_L
= - \alpha_3 - \alpha_4 - 2b_E N_R + 2 Q_E, \cr
\frac{2a - \epsilon}{g_s}
&= \alpha_I - 2 Q_E = 
\alpha_1 + \alpha_2 + 2 b_E N_L - 2 Q_E = 
- \alpha_3 - \alpha_4 - 2 b_E N_R.
\end{split}
\ee

\subsection{equivalence of \eqref{A10} and \eqref{A01} with \eqref{NekA10}
and \eqref{NekA01}}

\label{A1detail}

If we use the following relations,
\be
1 - \frac{Q_E}{\alpha_I - Q_E}
=\frac{\alpha_I - 2 Q_E}{\alpha_I - Q_E} = \frac{2a - \epsilon}{2a}, \qq
1 + \frac{Q_E}{\alpha_I - Q_E}=
\frac{\alpha_I}{\alpha_I - Q_E} = \frac{2a + \epsilon}{2a},
\ee
\be
\left\langle\!\!\!\left\langle
b_E \sum_{I=1}^{N_L} x_I
\right\rangle\!\!\!\right\rangle_{\!\! N_L}
= \frac{b_E N_L(b_E N_L - Q_E + \alpha_1)}{(\alpha_I - 2 Q_E)}
= \frac{(a-m_1)(a-m_2)}{g_s(2a-\epsilon)},
\ee
\be
\left\langle\!\!\!\left\langle
b_E \sum_{J=1}^{N_R} y_J  
\right\rangle\!\!\!\right\rangle_{\!\! N_R}
= \frac{b_E N_R( - \alpha_4 - b_E N_R + Q_E)}{\alpha_I}
= - \frac{(a+m_3)(a+m_4)}{g_s(2a+\epsilon)},
\ee
\be
\alpha_2 + \left( \frac{\alpha_I - 2 Q_E}{\alpha_I - Q_E} \right)
\left\langle\!\!\!\left\langle
b_E \sum_{I=1}^{N_L} x_I 
\right\rangle\!\!\!\right\rangle_{\!\! N_L}= \frac{(a+m_1)(a+m_2)}{2a g_s},
\ee
\be
\alpha_3 + \left( \frac{\alpha_I}{\alpha_I - Q_E} \right)
\left\langle\!\!\!\left\langle
b_E \sum_{J=1}^{N_R} y_J  
\right\rangle\!\!\!\right\rangle_{\!\! N_R}
= - \frac{(a-m_3)(a-m_4)}{2a g_s},
\ee
we can see that \eqref{A10} and \eqref{A01} reproduce
the Nekrasov function \eqref{NekA10}
and \eqref{NekA01} respectively.

\subsection{$\mathcal{A}_2$}

\label{A2detail}

The second order $q$-expansion coefficient $\mathcal{A}_2$
of \eqref{calA} is given by
\bel{A2expand}
\begin{split}
\mathcal{A}_2 &= \frac{1}{2}
\left\langle\!\!\! \left\langle
\left( \alpha_2 + b_E \sum_{I=1}^{N_L} x_I \right)^2
\right\rangle\!\!\! \right\rangle_{\!\! N_L}
\left\langle\!\!\! \left\langle
\left( b_E \sum_{J=1}^{N_R} y_J \right)^2
\right\rangle\!\!\! \right\rangle_{\!\! N_R} \cr
& +\frac{1}{2}
\left\langle\!\!\! \left\langle
\left( b_E \sum_{I=1}^{N_L} x_I \right)^2
\right\rangle\!\!\! \right\rangle_{\!\! N_L}
\left\langle\!\!\! \left\langle
\left( \alpha_3 + b_E \sum_{J=1}^{N_R} y_J \right)^2
\right\rangle\!\!\! \right\rangle_{\!\! N_R} \cr
& + \left\langle\!\!\! \left\langle
\left( b_E \sum_{I=1}^{N_L} x_I \right) 
\left( \alpha_2 + b_E \sum_{I'=1}^{N_L} x_{I'} \right)
\right\rangle\!\!\! \right\rangle_{\!\! N_L}
\left\langle\!\!\! \left\langle
\left( b_E \sum_{J=1}^{N_R} y_J \right)
\left( \alpha_3 + b_E \sum_{J'=1}^{N_R} y_{J'} \right)
\right\rangle\!\!\! \right\rangle_{\!\! N_R} \cr
& - \frac{1}{2}
\left\langle\!\!\! \left\langle
b_E \sum_{I=1}^{N_L} x_I^2 + \alpha_2  
\right\rangle\!\!\! \right\rangle_{\!\! N_L}
\left\langle\!\!\! \left\langle
b_E \sum_{J=1}^{N_R} y_J^2
\right\rangle\!\!\! \right\rangle_{\!\! N_R} \cr
& - \frac{1}{2}\left\langle\!\!\! \left\langle
b_E \sum_{I=1}^{N_L} x_I^2 
\right\rangle\!\!\! \right\rangle_{\!\! N_L}
\left\langle\!\!\! \left\langle
b_E \sum_{J=1}^{N_R} y_J^2 + \alpha_3 
\right\rangle\!\!\! \right\rangle_{\!\! N_R}.
\end{split}
\ee
By using
\be
 \left( b_E \sum_{I=1}^{N_L} x_I \right)^2 
= b_E^2 P_{(2)}^{(1/b_E^2)}(x)
+ \frac{2b_E^2}{1+b_E^2} P_{(1^2)}^{(1/b_E^2)}(x), 
\ee
\be
b_E \sum_{I=1}^{N_L} x_I^2 = b_E \, m_{(2)}(x) = b_E
\left( P_{(2)}^{(1/b_E^2)}(x) - \frac{2 b_E^2}{1 + b_E^2}
P_{(1^2)}^{(1/b_E^2)}(x) \right),
\ee
and similar relations for the right part,
we can convert the symmetric polynomials in \eqref{A2expand}
into the Jack polynomials.
Through \eqref{JackMMT}, the exact expressions for these moments are
obtained:
\bel{MMT2a}
\begin{split}
\left\langle\!\!\!\left\langle
 \left( b_E \sum_{I=1}^{N_L} x_I \right)^2 
\right\rangle\!\!\!\right\rangle_{\!\! N_L}
& = \frac{(a-m_1)(a-m_2)(a-m_1-\epsilon_1)(a-m_2-\epsilon_1)}
{\epsilon_1( \epsilon_1 - \epsilon_2)( 2a - \epsilon)(2a - \epsilon - \epsilon_1)} \cr
&- \frac{(a-m_1)(a-m_2)(a-m_1-\epsilon_2)(a-m_2-\epsilon_2)}
{\epsilon_2( \epsilon_1 - \epsilon_2)( 2a - \epsilon)(2a - \epsilon - \epsilon_2)},
\end{split}
\ee
\bel{MMT2b}
\begin{split}
\left\langle\!\!\!\left\langle
b_E \sum_{I=1}^{N_L} x_I^2
\right\rangle\!\!\!\right\rangle_{\!\! N_L} 
&= 
- \frac{(a-m_1)(a-m_2)(a-m_1-\epsilon_1)(a-m_2-\epsilon_1)}
{g_s ( \epsilon_1 - \epsilon_2)(2a - \epsilon)(2a - \epsilon - \epsilon_1)} \cr
&\ \ \ +\frac{(a-m_1)(a-m_2)(a-m_1 - \epsilon_2)(a-m_2-\epsilon_2)}
{g_s ( \epsilon_1 - \epsilon_2)(2a - \epsilon)(2a- \epsilon - \epsilon_2)}.
\end{split}
\ee
By substituting these formula into \eqref{A2expand},
the expansion coefficient $\mathcal{A}_2$ is obtained. 
Since the result is rather lengthy, we do not write it here.
But we have checked that it exactly coincides with the
sum of the five Nekrasov functions:
\bel{A2to5}
\mathcal{A}_2 = 
\mathcal{A}_{(2),(0)}^{\mathrm{Nek}}
+ \mathcal{A}_{(1^2),(0)}^{\mathrm{Nek}}
+ \mathcal{A}_{(1),(1)}^{\mathrm{Nek}}
+ \mathcal{A}_{(0),(1^2)}^{\mathrm{Nek}}
+ \mathcal{A}_{(0),(2)}^{\mathrm{Nek}}.
\ee
An explicit form of these $\mathcal{A}_{Y_1,Y_2}^{\mathrm{Nek}}$
can be found, for example, in \cite{MMM0907}.

\subsubsection{qualitative explanation on rearrangements}

Unfortunately, we do not have a systematic algorithm for rearrangements of
the exact expression into the Nekrasov functions \eqref{A2five} such that
each term has a factorized form \eqref{PolyMM}.
But a general pattern can be easily read out from \eqref{A2expand}.
Note that the mass dependence of
the moments \eqref{MMT2a} and \eqref{MMT2b} are
given by certain symmetric polynomial of $a-m_1$ and $a-m_2$.
They do not contain $a+m_1$ and $a+m_2$. Likewise, on the 
right part, the moments depend on $a+m_3$ and $a+m_4$ symmetrically.
The sign flip of mass is caused by mixture with the 
constants $\alpha_2 = (m_1 +m_2)/g_s$
(or $\alpha_3 = (m_3+m_4)/g_s$ for the right part).
For example, let us consider the first term in \eqref{A2expand}:
\bel{A2FstT}
\frac{1}{2}
\left\langle\!\!\! \left\langle
\left( \alpha_2 + b_E \sum_{I=1}^{N_L} x_I \right)^2
\right\rangle\!\!\! \right\rangle_{\!\! N_L}
\left\langle\!\!\! \left\langle
\left( b_E \sum_{J=1}^{N_R} y_J \right)^2
\right\rangle\!\!\! \right\rangle_{\!\! N_R}.
\ee
The term on the right part contains the Jack polynomial $P_{(2)}^{(1/b_E^2)}(y)$:
\be
\left( b_E \sum_{J=1}^{N_R} y_J \right)^2
= b_E^2 P_{(2)}^{(1/b_E^2)}(y) + \dotsm,
\ee
thus this ``dominant'' part gives
\bel{EqA24}
b_E \times
\Bigl\langle\!\!\Bigl\langle
b_E P_{(2)}^{(1/b_E^2)}(y)
\Bigr\rangle\!\!\Bigr\rangle_{\! N_R}
= - \frac{(a+m_3)(a+m_4)(a+m_3+\epsilon_2)(a+m_4+\epsilon_2)}
{\epsilon_2 ( \epsilon_1 - \epsilon_2) ( 2 a + \epsilon)
(2a + \epsilon + \epsilon_2)}.
\ee
In the left part, a quite non-trivial portion
$M_{(2),(0)}(x)$ of the total terms gives a ``dominant''
contribution:
\be
\begin{split}
\frac{1}{2}
\left( \alpha_2 + b_E \sum_{I=1}^{N_L} x_I \right)^2
&= \frac{1}{2} \alpha_2^2 + \alpha_2 b_E 
\sum_{I=1}^{N_L} x_I 
+ \frac{1}{2} \left( b_E \sum_{I=1}^{N_L} x_I \right)^2 \cr
&= M_{(2),(0)}(x) + \mbox{sub-dominant terms},
\end{split}
\ee
where
\bel{M20x}
\begin{split}
M_{(2),(0)}(x)&:= 
\frac{1}{2} \alpha_2^2 - \frac{\alpha_2}{2b_E}
+ \frac{(2a - \epsilon_1) \alpha_2}{(2a+\epsilon_2)}
\left( b_E\, 
P_{(1)}^{(1/b_E^2)}(x) \right)\cr
&-
\frac{(2a-\epsilon - \epsilon_2)}{\epsilon_2
(\epsilon_1 - \epsilon_2)(2a+\epsilon_2)}
\left(
2 b_E^2\, P_{(1^2)}^{(1/b_E^2)}(x) \right) \cr
&+ \frac{(\epsilon_1 + \epsilon_2)(2a - \epsilon_1 + \epsilon_2)
(2a - \epsilon - \epsilon_2)}{4ag_s(2a+\epsilon_2)}
\left(
b_E P_{(2)}^{(1/b_E^2)} (x) \right).
\end{split}
\ee
Here, we have determined this polynomial 
by requiring the linear factorization of the moment (as in \eqref{EqA24})
and also by
the assumption on  mass dependence only on $a+m_1$ and $a+m_2$ symmetrically.
Indeed, while the form of the polynomial
\eqref{M20x} is messy, its moment takes a quite simple form:
\be
\Bigl\langle\!\! \Bigl\langle
M_{(2),(0)}(x) 
\Bigr\rangle\!\! \Bigr\rangle_{\! N_L}
=- \frac{(a+m_1)(a+m_2)(a+m_1+ \epsilon_2)(a+m_2+ \epsilon_2)}
{4\, \epsilon_1 \epsilon_2 \, a(2a + \epsilon_2)}.
\ee
Hence, seen at the level of moment it is essentially dual of the Jack polynomial 
$P_{(2)}^{(1/b_E^2)}(x)$ such that
the sign of $a$ is reversed.

Now by setting 
\bel{M20y}
\widetilde{M}_{(2),(0)}(y):= b_E^2 P_{(2)}^{(1/b_E^2)}(y),
\ee
we can see that the first term \eqref{A2FstT} contains
the portion
\bel{M20xy}
\Bigl\langle\!\!\Bigl\langle
M_{(2),(0)}(x) 
\Bigr\rangle\!\! \Bigr\rangle_{\! N_L}
\Bigl\langle\!\!\Bigl\langle
\widetilde{M}_{(2),(0)}(y) 
\Bigr\rangle\!\! \Bigr\rangle_{\! N_R}
= \frac{\prod_{i=1}^4 ( a + m_i )( a + m_i + \epsilon_2)}
{4 \, \epsilon_1 \epsilon_2^2 \, a (2a+\epsilon_2)
(2a + \epsilon)
(2a + \epsilon + \epsilon_2)},
\ee
which is exactly equal to
$\mathcal{A}_{(2),(0)}^{\mathrm{Nek}}$. 

Note that at $b_E=1$, the symmetric polynomial $M_{(2),(0)}(x)$
\eqref{M20x} turns into the form
\be
M_{(2),(0)}(x) =  \frac{1}{2} \alpha_2^2 - \frac{1}{2} \alpha_2
+ \alpha_2 P_{(1)}^{(1)}(x) +  P_{(1^2)}^{(1)}(x).
\ee
It has quite simple coefficients and the dominant term
is $P_{(1^2)}^{(1)}(x) = m_{(1^2)}(x)$,
the monomial symmetric polynomial characterized by
the conjugate partition $Y_1'=(1^2)$ of the partition $Y_1 = (2)$.

\subsubsection{polynomials $M_{Y_1,Y_2}(x)$ and $\widetilde{M}_{Y_1,Y_2}(y)$}

Here we write the polynomials $M_{Y_1,Y_2}(x)$ 
and their pairs $\widetilde{M}_{Y_1,Y_2}(y)$ \eqref{PolyMM} with $|Y_1|+|Y_2|=2$.

For $Y_1=(2)$ and $Y_2=(0)$, they are given by \eqref{M20x} and \eqref{M20y}
respectively. While for $Y_1=(1^2)$ and $Y_2=(0)$,
\be
\begin{split}
M_{(1^2),(0)}(x)
&=\frac{1}{2} \alpha_2^2 + \frac{1}{2} b_E\, \alpha_2
+ \left(\frac{2a - \epsilon_2}{2a + \epsilon_1} \right) \alpha_2\, 
b_E P_{(1)}^{(1/b_E^2)}(x) , \cr
& - \left(
\frac{ \epsilon
(2a + \epsilon_1 - \epsilon_2)(2a- \epsilon - \epsilon_1)}
{2(\epsilon_1 - \epsilon_2)
a(2a+ \epsilon_1)} \right) b_E^2 P_{(1^2)}^{(1/b_E^2)}(x) 
\cr
& + \left( \frac{2a - \epsilon - \epsilon_2}{2a + \epsilon_1} \right)
b_E^2P_{(2)}^{(1/b_E^2)}(x), \cr
\widetilde{M}_{(1^2),(0)}(y)&= \frac{2 b_E^2}{1+b_E^2}
P_{(1^2)}^{(1/b_E^2)}(y).
\end{split}
\ee
Also, for $Y_1=(1)$ and $Y_2=(1)$,
\be
\begin{split}
M_{(1),(1)}(x)
&= \alpha_2 g_s b_E P_{(1)}^{(1/b_E^2)}(x)
- \frac{g_s \epsilon_2 ( 2a - \epsilon - \epsilon_1)}
{(\epsilon_1-\epsilon_2)
(2a - \epsilon_1)}
\, 2 b_E^2\, P_{(1^2)}^{(1/b_E^2)}(x) \cr
&+ \epsilon_1 
\left(\frac{2a - \epsilon - \epsilon_2}{2 a - \epsilon_2} \right) b_E
P_{(2)}^{(1/b_E^2)}(x).
\end{split}
\ee
The rest can be obtained easily from these formula by the exchange
$M_{Y_1,Y_2}(x) \longleftrightarrow \widetilde{M}_{Y_2,Y_1}(y)$
along with the replacement of parameters \eqref{GTparity}.

The moments for $M_{Y_1,Y_2}(x)$ are given by
\be
\begin{split}
\Bigl\langle\!\!\Bigl\langle
M_{(1^2),(0)}(x)
\Bigr\rangle\!\!\Bigr\rangle_{\! N_L}
&= \frac{(a+m_1)(a+m_2)(a+m_1 + \epsilon_1)(a+m_2+\epsilon_1)}
{4 g_s^2 a(2a+ \epsilon_1)}, \cr
\Bigl\langle\!\!\Bigl\langle
M_{(1),(1)}(x)
\Bigr\rangle\!\!\Bigr\rangle_{\! N_L}
&= \frac{(a+m_1)(a+m_2)(a-m_1)(a-m_2)}
{g_s (2a - \epsilon_1)(2a-\epsilon_2)}.
\end{split}
\ee

Note that at $b_E=1$,
\be
\begin{split}
M_{(1^2),(0)}(x)
&= \frac{1}{2} \alpha_2^2 + \frac{1}{2} \alpha_2
+ \alpha_2 P_{(1)}^{(1)}(x) + P_{(2)}^{(1)}(x), \cr
M_{(1),(1)}(x)
&= g_s \alpha_2 P_{(1)}^{(1)}(x) + g_s P_{(1^2)}^{(1)}(x)
+ g_s P_{(2)}^{(1)}(x).
\end{split}
\ee

\section{Free Energy from Selberg Integral}

In this section, we briefly discuss 
the contribution to
the fee energy of the perturbed
double-Selberg model  \eqref{PDS}, i.e., $Z_{(\mathrm{Selberg})^2}$:
\be
Z_{(\mathrm{Selberg})^2} = \exp\left( \mathcal{F}_{L}
+ \mathcal{F}_{R} \right),
\ee
where
\be
\mathcal{F}_{L}= \log S_{N_L}(1 + b_E \alpha_1, 1 + b_E \alpha_2, b_E^2), \qq
\mathcal{F}_{R} = \log S_{N_R}(1 + b_E \alpha_4, 1 + b_E \alpha_3, b_E^2).
\ee
Since we have an exact expression of the Selberg integral $S_N(\beta_1, \beta_2, \gamma)$
\eqref{SELB}, an integral representation of the free energy is easily obtained:
\bel{FreeE}
\mathcal{F}_{L}
= \int_0^{\infty} \frac{\de t}{t}
\ex^{-t} \left(\frac{(1-\ex^{-N_L b_E^2 t})
\widetilde{\mathcal{G}}_L(t)}{(1-\ex^{-t})(1 - \ex^{-b_E^2 t})}
- N_L \frac{(1+\ex^{-b_E^2 t})}{(1 - \ex^{-t})} - N_L
\right),
\ee
where
\be
\begin{split}
\widetilde{\mathcal{G}}_L(t)
&= \exp\left(-b_E^2 t\right) + 
\exp\left(-b_E \alpha_1 t \right)
+ \exp\left(-b_E \alpha_2 t \right)  \cr
&- \exp\Bigl(-\bigl\{ 1+b_E(\alpha_1+\alpha_2) + (N_L-1) b_E^2 \bigr\} t\Bigr),
\end{split}
\ee
and similar expression for $\mathcal{F}_{R}$.

\subsection{$g_s$-expansion of the free energy}

For simplicity, we assume the deformation parameter $b_E$ real and positive.
Then, the free energy \eqref{FreeE}
can be rewritten by using the parameters of the gauge theory.
Through the relation \eqref{0d4d}, we have
\bel{FLPert}
\mathcal{F}_{L} = \int_0^{\infty}
\frac{\de t}{t} \, \ex^{-(g_s/b_E) t}
\left[ \frac{(1 - \ex^{-(a-m_2) t} ) \mathcal{G}_L(t)}
{(1-\ex^{-g_s b_E t})(1 - \ex^{-(g_s/b_E)t})}
- \frac{(a-m_2)}{b_E g_s}
\left( \frac{(1+\ex^{-g_s b_E t})}{(1 - \ex^{-(g_s/b_E)t})} + 1
\right) \right],
\ee
where
\be
\mathcal{G}_L(t)
= \ex^{-g_s b_E t} + \ex^{-(m_2-m_1+g_s Q_E)t}
+ \ex^{-(m_2+m_1)t} - \ex^{-(a+m_2)t}.
\ee
Let us consider the
$g_s$-expansion of \eqref{FLPert}:
\be \label{FEgsexp}
\mathcal{F}_{L} 
= \sum_{k=0}^{\infty} g_s^{k-2} \mathcal{F}_{L}^{(k/2)}
= \frac{1}{g_s^2} \mathcal{F}_{L}^{(0)}
+ \frac{1}{g_s} \mathcal{F}_{L}^{(1/2)}
+ \mathcal{F}_{L}^{(1)} + \dotsm.
\ee
For example, the planar part is given by
\be
\begin{split}
\mathcal{F}_{L}^{(0)}
&= \frac{1}{2} \sum_{i=1}^2 \Bigl\{ ( a + m_i)^2 \log (a + m_i)
+ ( a - m_i)^2 \log(a - m_i) \Bigr\} \cr
& - 2 a^2 \log (2a)
 - \frac{1}{2} (m_2 + m_1)^2 \log (m_2 + m_1)
- \frac{1}{2} ( m_2 - m_1)^2 \log (m_2 - m_1).
\end{split}
\ee
Combining with $\mathcal{F}_R^{(0)}$, we obtain
a perturbative part of the prepotential for the
gauge theory.


\end{document}